\documentclass[12pt,a4paper,oneside]{book}

\usepackage[latin1]{inputenc}
\usepackage{textcomp}
\usepackage[T1]{fontenc}
\usepackage[italian]{babel}
\usepackage{graphicx}
\usepackage{lscape}
\usepackage{latexsym}
\usepackage{amssymb}
\usepackage{fancyhdr}

\newcommand{\enfatizza}{\bfseries}
\newcommand{\parla}{\em}
\newcommand{\sottolinea}{\em}
\newcommand{\parolestrane}{\em}
\newtheorem{fig}{Figura}
\newtheorem{Teorema}{Teorema:}

\begin{document}

\titlepage

\begin{center}

\begin{LARGE}

UNIVERSITA' DEGLI STUDI DI TORINO \\
FACOLTA' DI SCIENZE M.F.N. \\
CORSO DI LAUREA IN FISICA  \\

\end{LARGE}

\vspace{1cm} 

TESI DI LAUREA

\vspace{2cm}

\begin{bfseries}

REALIZZAZIONE DI UN ESPERIMENTO
INNOVATIVO CON DOPPIA FENDITURA 
UTILIZZANDO COPPIE CORRELATE DI
FOTONI.

\end{bfseries}

\vspace{2cm}

\begin{minipage}[b]{6cm}

Relatore:\\
Pres. E. Predazzi

\end{minipage}
\ \hspace{5mm} \hspace{5mm} \ 
\begin{minipage}[b]{6cm}

Relatori esterni:\\
ing. Brida e dott. Genovese

\end{minipage}

\vspace{4cm}

Tesista:\\
Falzetta Giuseppe

\vspace{29mm}

Torino, gennaio 2003

\end{center}

\clearpage\nonumber

\begin{scriptsize}

A mio  {\bfseries padre},                             \\
che ha tanto desiderato                               \\
che io prendessi una laurea                           \\
(una qualsiasi,{\parla `anche in fisica va bene'})

e a mia {\bfseries madre},                            \\
che pur avendo conseguito solo                        \\
la quinta elementare ha sempre fatto                  \\
finta di capire i miei studi.                         \\

\end{scriptsize}

\clearpage

\begin{center}
{\bfseries RINGRAZIAMENTI}\\
\end{center}
Voglio porgere, in primo luogo, i miei più sentiti ringraziamenti                \\
al mio relatore, il pres. Enrico Predazzi, per i suoi preziosi consigli          \\
fondamentali non solo per il mio lavoro di tesi, ma anche per la mia             \\
formazione culturale.                                                                      \\
Al dott. Marco Genovese e all'ing. Giorgio Brida dell'Istituto Elettrotecnico     \\
Nazionale Galileo Ferraris vanno i miei ringraziamenti per l'opportunità         \\
che mi è stata concessa di realizzare una tesi estremamente  innovativa in       \\
un ambiente molto stimolante per la ricerca.                                     \\
Le persone che mi hanno aiutato e che vorrei ringraziare  sono molte,            \\
ognuno ha fornito un contributo prezioso, ma citarli tutti non è possibile.      \\
Ad Emanuele Cagliero, Marco Gramegna e Gianna Panfilo un grazie per              \\
tutte le preziose nozioni informatiche che mi hanno permesso di                  \\
trovare la soluzione ai problemi che affliggono chi come me conosce poco         \\
il mondo del computer.\\

\vspace{4cm}

Giuseppe Falzetta

\clearpage

\tableofcontents

\chapter{Introduzione}

\begin{figure}[h]
\begin{minipage}[b]{6cm}
   \centering
   \includegraphics[width=5cm]{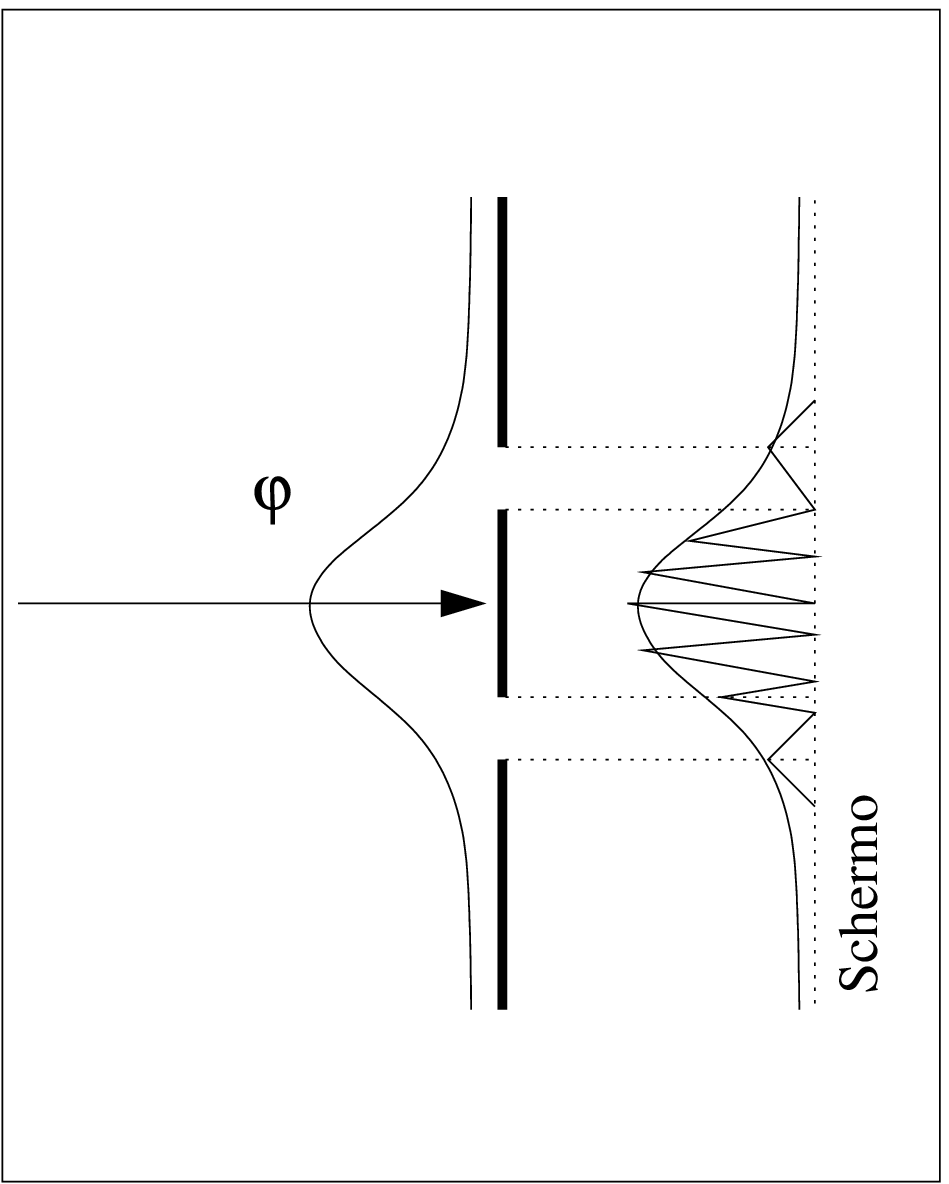}
         \begin{fig}\label{onda}
Figura di diffrazione e di interferenza 
generata dal passaggio 
attraverso una doppia fenditura 
di un fotone descritto
da una funzione d'onda~$\varphi$ 
sufficientemente larga da avere valore non nullo 
in corrispondenza delle due aperture di 
una doppia fenditura.
         \end{fig}
\end{minipage}
\ \hspace{5mm} \hspace{5mm} \ 
\begin{minipage}[b]{6cm}
   \centering
   \includegraphics[width=5cm]{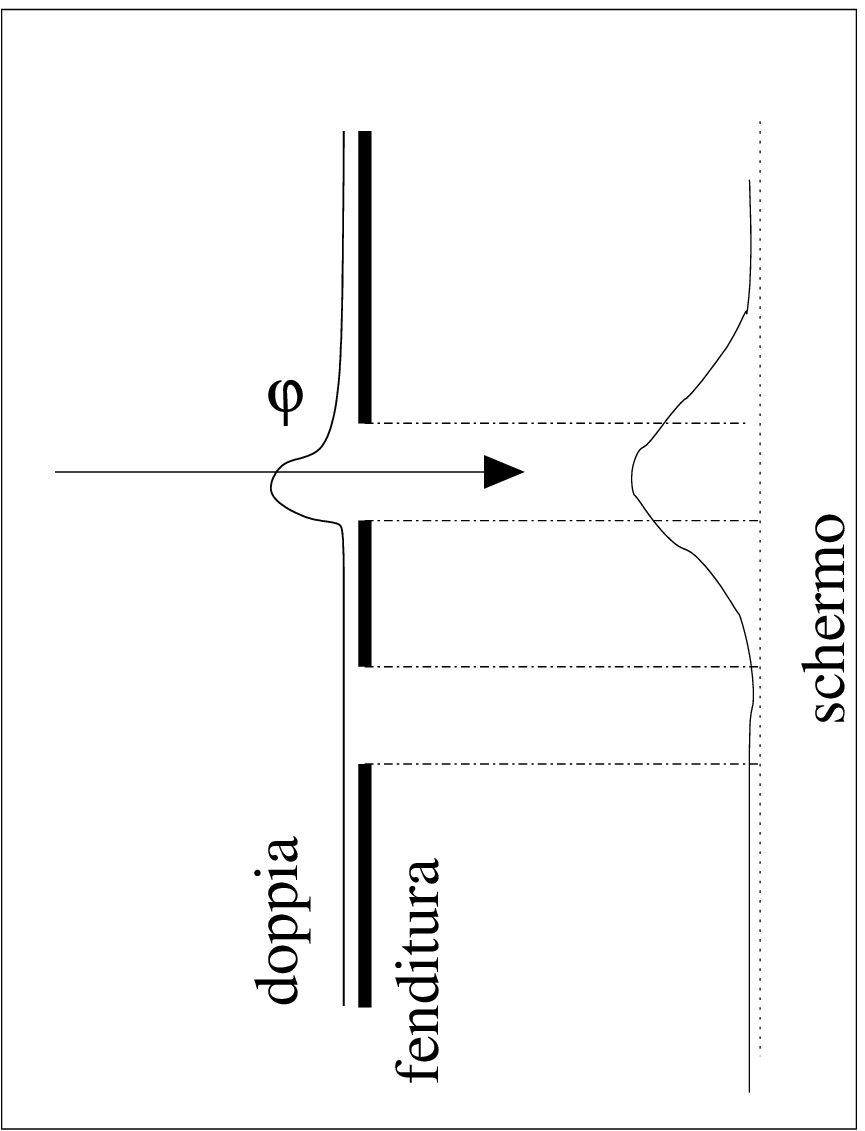}
         \begin{fig}\label{particella}
Figura di diffrazione generata dal passaggio 
attraverso una specifica apertura 
di una doppia fenditura 
di un fotone descritto
da una funzione d'onda~$\varphi$ stretta.\\
E' da notare l'assenza del fenomeno 
dell'interferenza.
         \end{fig}
\end{minipage}
\end{figure}

In meccanica classica i fenomeni fisici di tipo
ondulatorio o corpuscolare sono 
chiaramente distinti:
ad esempio, la pressione di un gas
è concepita in termini della teoria
cinetica molecolare,
mentre per la propagazione del suono è 
efficace un'interpretazione ondulatoria.
In meccanica quantistica, invece,
tale distinzione viene meno,
un sistema fisico è descritto
in maniera completa da una funzione d'onda 
il cui modulo quadro dà una 
distribuzione di probabilità
di posizione.
Solo all'atto della misura il
sistema `collassa' in una posizione
definita (entro i limiti posti
dal principio di indeterminazione).
Ad esempio in meccanica quantistica
la radiazione elettromagnetica
è rivelata come `quanti'
di energia $h\nu$,
ma la propagazione è descritta
dalla meccanica ondulatoria e si
hanno fenomeni di interferenza,
diffrazione, etc.
La stessa situazione si presenta 
per ogni `particella' 
(elettrone, protone, etc.)
a cui si associa una lunghezza d'onda
$\lambda=h/p$ ove $p$ è l'impulso.
Questo aspetto tipico della teoria dei 
quanti è espresso dal principio
di complementarità:
{\parla `Gli aspetti ondulatorio e corpuscolare
sono complementari, ed esistono solamente
come potenzialità;
un esperimento può convertire questa potenzialità
in un fenomeno osservabile, 
ma l'osservazione di uno dei due 
aspetti esclude l'altro'}~\cite{complementarità}.\\
Gli esperimenti basati sull'uso
di una doppia fenditura consentono uno studio
di tale principio mettendo in rilievo 
gli aspetti caratteristici della 
meccanica quantistica riguardo tali fenomeni.\\
Si consideri, ad esempio, il caso in cui
un fotone (o un'altra `particella') 
venga inviato contro una
doppia fenditura;
se la funzione d'onda $\varphi$ 
che lo descrive è sufficientemente larga\footnote{
L'aggettivo `larga' è usato 
in contrapposizione al termine `stretta'
con cui si intende una funzione d'onda la cui distribuzione 
di probabilità spaziale sia tale da essere non trascurabile
solo in corrispondenza di una delle due fenditure.} 
cioè tale 
da avere valore non nullo in corrispondenza
delle due aperture~(fig.~\ref{onda})
su uno schermo, posto dopo la doppia fenditura, 
si possono osservare (raccogliendo molti fotoni)
delle frange di interferenza,
fenomeno tipico delle onde,
le quali sono diffuse nello 
spazio.
Non si può quindi attribuire al fotone
una traiettoria specifica,
esso è descritto da una funzione d'onda
non localizzata e quindi non si
può affermare `quale fenditura'
esso attraversi.
L'interferenza tra le componenti
della funzione d'onda corrispondente
all'attraversamento di una o l'altra 
fenditura dà origine alla distribuzione
di probabilità di rilevazione del
fotone sullo schermo.\\
Se, invece, la $\varphi$ è stretta~(fig.~\ref{particella}),
e cioè ha un valore diverso da zero 
solo in prossimità di una delle due 
aperture, in maniera che si possa identificare
con certezza attraverso quale determinata apertura
esso sia passato 
(comportandosi come una particella),
non si ha interferenza.\\
Esistono anche situazioni intermedie in cui è possibile
una parziale identificazione della traiettoria dove
l'indistinguibilità del percorso\footnote{
Ved. ref.~\cite{auletta} per una definizione di~D.
}
$D$
diminuisce la visibilità~$V$ dell'interferenza  secondo
la relazione:

\begin{equation}\label{percorsointerferenza}
D^2+V^2 \leq 1
\end{equation}

\clearpage

Numerosi esperimenti sono stati realizzati col
fine di investigare ulteriormente
il principio di complementarità e sono basati
su studi della figura 
di interferenza prodotta da fotoni,
elettroni, protoni, etc,
che passano attraverso una doppia fenditura.
Tali studi rappresentano
delle varianti rispetto all'idea originaria
realizzata da Young, agli inizi del~XIX
secolo, sulla luce.

\begin{figure}[h]
\begin{center}
\includegraphics[width=14cm]{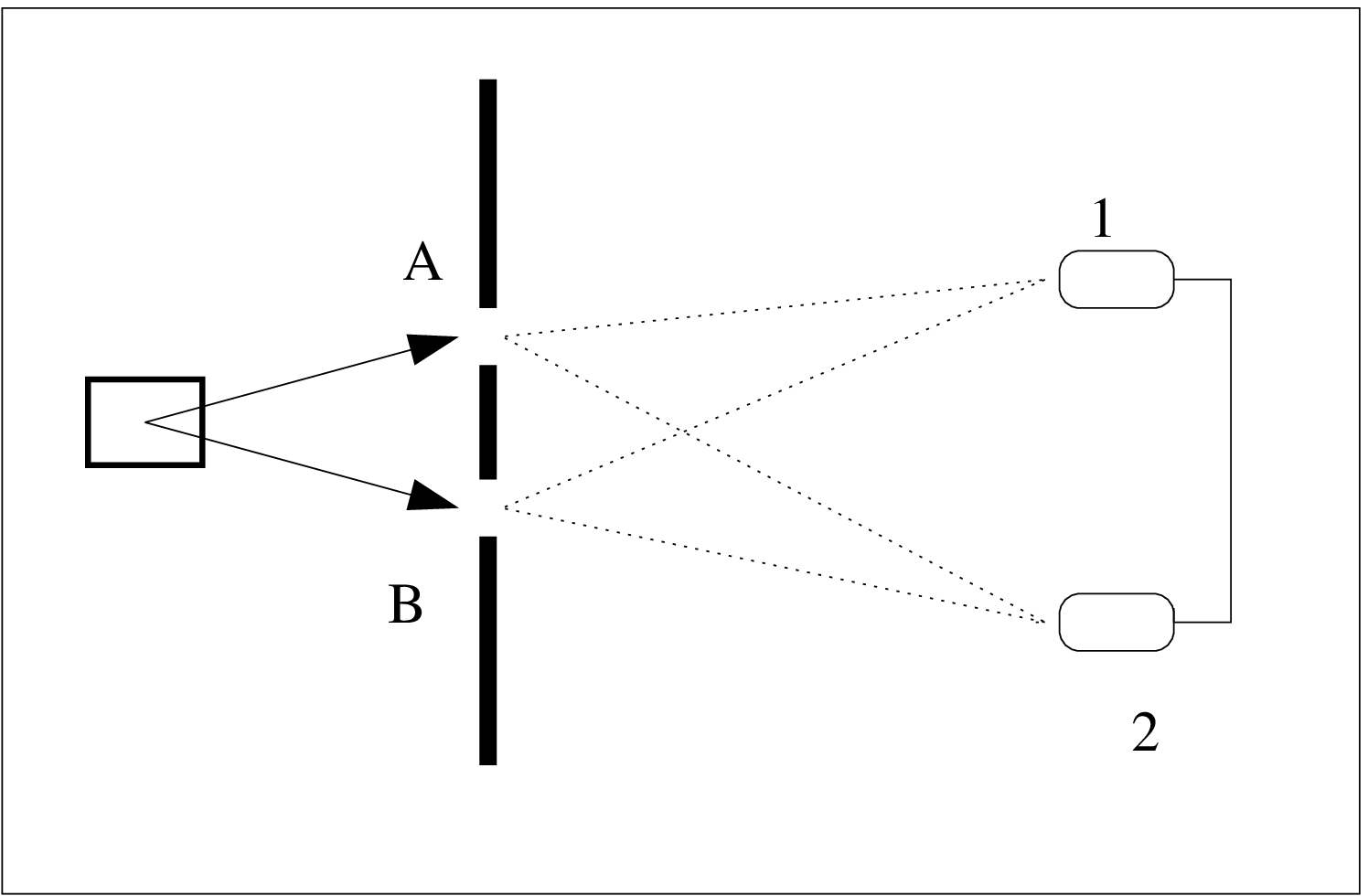}
\end{center}
       \begin{fig}\label{schemadetector}
Schema della configurazione sperimentale
adottata per la nostra esperienza, ove per convenzione 
si indicano con le lettere~A e~B le aperture della 
doppia fenditura e con~1 e~2 i fotorivelatori.
       \end{fig}
\end{figure}

Recentemente sono stati realizzati lavori,
\cite{2}-\cite{5},
che verranno descritti nel capitolo~\ref{esperimenticondoppiafenditura},
in cui si utilizzano fotoni prodotti
per fluorescenza parametrica,
fenomeno quantistico
senza analogo 
classico e di cui si forniscono alcuni cenni
nel capitolo successivo.\\
L'esperimento oggetto della presente tesi
rientra in questa categoria,
ma a differenza delle esperienze già svolte,
nel nostro caso entrambi i fasci 
({\parolestrane signal} ed {\parolestrane idler}) 
prodotti per fluorescenza parametrica
vengono indirizzati 
contro una doppia fenditura,
la quale è stata posizionata in modo 
tale che ogni fascio attraversi una 
specifica apertura.
I fotoni utilizzati sono descritti da una funzione
d'onda~$\varphi$ stretta,
cioè con valore non nullo solo in una piccola regione spaziale.
Tale caratteristica consente di
non avere interferenza 
a livello di singolo fotone (II ordine),
ma solo al quarto ordine (coincidenze) 
su uno schermo
posto a grande distanza da questa.
Infatti, mentre il cammino di singolo fotone
è perfettamente identificato e
la fenditura che esso attraversa è nota,
a livello delle coincidenze tra i due fotoni
della coppia~(fig.~\ref{schemadetector}) 
non è possibile identificare 
se il fotone rivelato da un dispositivo~(1 o 2)
abbia attraversato una delle due fenditure~(A o B) 
o l'altra.\\
Tale esperimento rappresenta, quindi,
un ulteriore emblematico esempio
del legame tra conoscenza del percorso
ed interferenza.

In tale esperienza, inoltre, è stata realizzata
la proposta di 
P.Ghose~\cite{propostaesperimentoghose}
volta,
mediante l'uso del formalismo sviluppato da
Kemmer - Duffin - Harishchandra~\cite{formalismoghose},
ad un confronto
tra la meccanica quantistica standard
(nel seguito SQM)
e la teoria di de~Broglie-Bhom 
(dBB),
una delle più significative
teorie a variabili nascoste non-locale
che viene descritta nel capitolo~\ref{dBB}.\\
I nostri risultati sono in accordo con la SQM, 
ma contraddicono le previsioni teoriche di
Ghose per la dBB di 8 deviazioni standard.
Si allega in appendice~\ref{pubblicazione}
l'articolo pubblicato, ref.\cite{articolonostro1},
in cui sono presenti tali risultati.

In appendice~\ref{pre-preprint}, invece, è riportato
l'articolo (che sarà presente in forma di pre-print
tra breve tempo) che contiene lo studio completo 
della figura di interferenza.

Infine, in appendice~\ref{pubblicazionenewscientist}
si allega l'articolo di divulgazione scientifica 
ref.\cite{articolonewscientist} in cui sono  
discusse le nostre prime conclusioni.

\chapter{Entanglement e f{}luorescenza parametrica}\label{entanglement}

\section{Stati quantistici entangled}

Col termine di {\parolestrane entangled} 
si definiscono stati quantistici,
di due o più particelle,
per le quali la funzione d'onda non è 
fattorizzabile nel prodotto di funzioni 
d'onda di particelle singole.

Per darne una definizione più generale
si introduce la
{\enfatizza matrice densità} 
di uno stato~$|\psi \rangle$ definita come:

\begin{equation}
\rho_{\psi}=|\psi \rangle \langle\psi|
\end{equation}

Per un sistema composto da due sottosistemi A e B 
si può scrivere:

\begin{equation}\label{rho}
\rho=|\psi \rangle _{AB}{}_{AB}\langle\psi| 
\end{equation}

\begin{equation}\label{stato}
|\psi \rangle _{AB}=\sum_{j,\mu}a_{j\mu}|j \rangle _{A}\otimes|\mu \rangle _{B}
\end{equation}

\clearpage

Usando la 
{\enfatizza decomposizione di Schmidt},  
lo stato~\ref{stato} assume la seguente forma:

\begin{equation}\label{entangledgenerale}
|\psi \rangle _{AB}=\sum_{i=1}^{n} c_i|\psi^i \rangle _A|\phi^i \rangle _B
\end{equation}

Dove $|\phi^i \rangle _{B}$ e $|\psi^i \rangle _{A}$
sono stati ortonormali che diagonalizzano
contemporaneamente la matrice densità~\ref{rho}:

\begin{equation}
|\phi^i \rangle _{B}=\sum_{\mu}b_{i\mu}|\mu \rangle _{B}
\end{equation}

\begin{equation}
|\psi^i \rangle _{A}=\sum_{\nu}d_{i\nu}|\nu \rangle _{A}
\end{equation}
  
Si può quindi dare la seguente 
definizione:
{\parla ` Uno stato del sistema
bipartito è entangled se e solo se la sua decomposizione
di Schmidt ha più di un solo termine~'},
in base alla quale, utilizzando
la~\ref{entangledgenerale}
segue che un generico stato quantistico
{\enfatizza entangled} a 2 componenti
può essere scritto nel seguente modo:\\

\begin{equation}
|\psi \rangle =c_1|\psi^1 \rangle _A|\phi^1 \rangle _B+%
c_2|\psi^2 \rangle _A|\phi^2\rangle _B
\end{equation}

I pedici individuano gli spazi
di Hilbert in cui sono definiti i ket,
gli apici, i vettori.\\
Un esempio 
è fornito dallo stato a due fotoni:

\begin{equation}
|\Psi \rangle =a|\Psi_1^{\alpha}\Psi_2^{\beta} \rangle +b|\Psi_1^{\gamma}\Psi_2^{\delta}\rangle 
\end{equation}

Dove gli apici denotano le polarizzazioni, 
mentre gli indici 1 e 2 la direzione per il fotone.\\

\section{Fluorescenza parametrica}

\begin{figure}[h]
\begin{minipage}[b]{6cm}
   \centering
   \includegraphics[width=6cm]{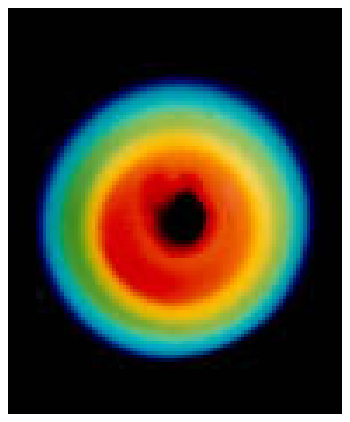}
       \begin{fig}\label{pdcItipo}
Fotografia della fluorescenza parametrica di tipo~I
(utilizzata per il nostro esperimento).
       \end{fig}
\end{minipage}
\ \hspace{5mm} \hspace{5mm} \
\begin{minipage}[b]{6cm}
   \centering
   \includegraphics[width=6cm]{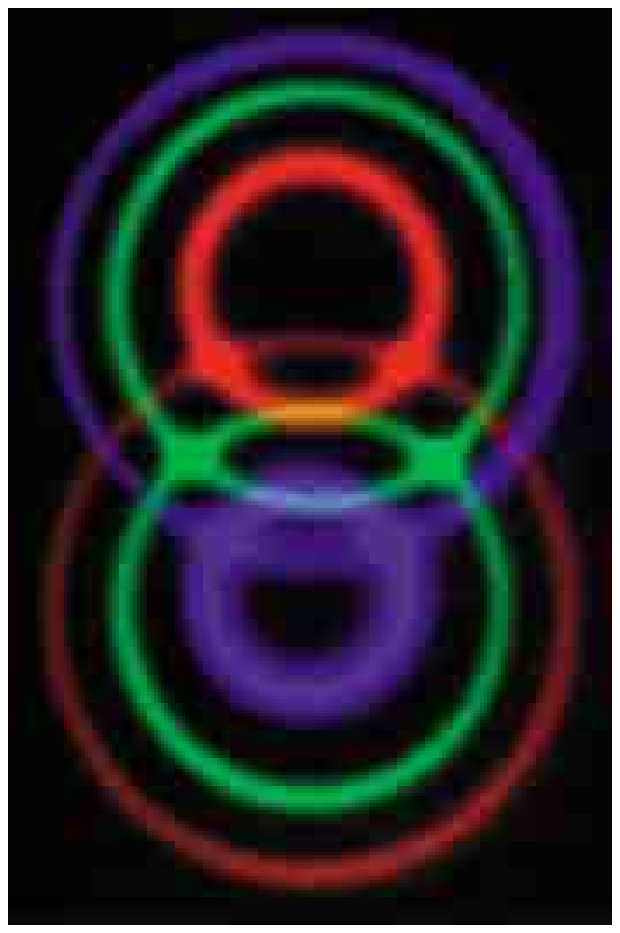}
       \begin{fig}\label{pdcIItipo}
Fotografia della fluorescenza parametrica di tipo~II.
       \end{fig}
\end{minipage}
\end{figure}

Fin dagli anni '70 stati {\parolestrane entangled} a due fotoni 
vengono prodotti mediante la
{\parolestrane parametric down-conversion}~(nel
seguito PDC), fenomeno di ottica non lineare che avviene tramite
interazione di una radiazione laser con un opportuno dielettrico,
il quale è caratterizzato da  coefficienti di 
suscettività elettrica con valori  
non trascurabili agli ordini superiori al primo.\\
Tipicamente si tratta di cristalli anisotropi,
che possono avere uno o due assi di simmetria:
nel primo caso si ha la PDC di  tipo I
(fig.~\ref{pdcItipo}),
nel secondo quella di  tipo II
(fig.~\ref{pdcIItipo}).

\clearpage

Più in dettaglio, la PDC avviene allorquando una
radiazione monocromatica  di frequenza  $\omega_3$
(normalmente detta "pompa") 
incide con un'opportuna polarizzazione 
su un mezzo non lineare,
tipicamente un cristallo 
uniassico, come onda straordinaria.\\
Si dimostra che, per effetto del valore di 
$\chi_{ijk}^{(2)}$,
tensore suscettività elettrica
di second'ordine,
diverso da zero esiste una probabilità non nulla, 
proporzionale al quadrato di
$\chi_{ijk}^{(2)}$ , 
che un fotone  
$\omega_3$
incidente `decada' in due fotoni  
$\omega_1$
e
$\omega_2$
a frequenza più bassa, 
da cui il temine "down-conversion", 
soddisfacendo la richiesta della conservazione 
dell'energia e della quantità di moto,
ossia:

\begin{eqnarray}\label{vincolo1}
\omega_1+\omega_2=\omega_3  \label{omega}
\end{eqnarray}

\begin{eqnarray}
\overrightarrow{k_1}+\overrightarrow{k_2}=\overrightarrow{k_3} \label{pm}
\end{eqnarray}

\begin{figure}[h]
\begin{minipage}[b]{6cm}
   \centering
   \includegraphics[width=6cm]{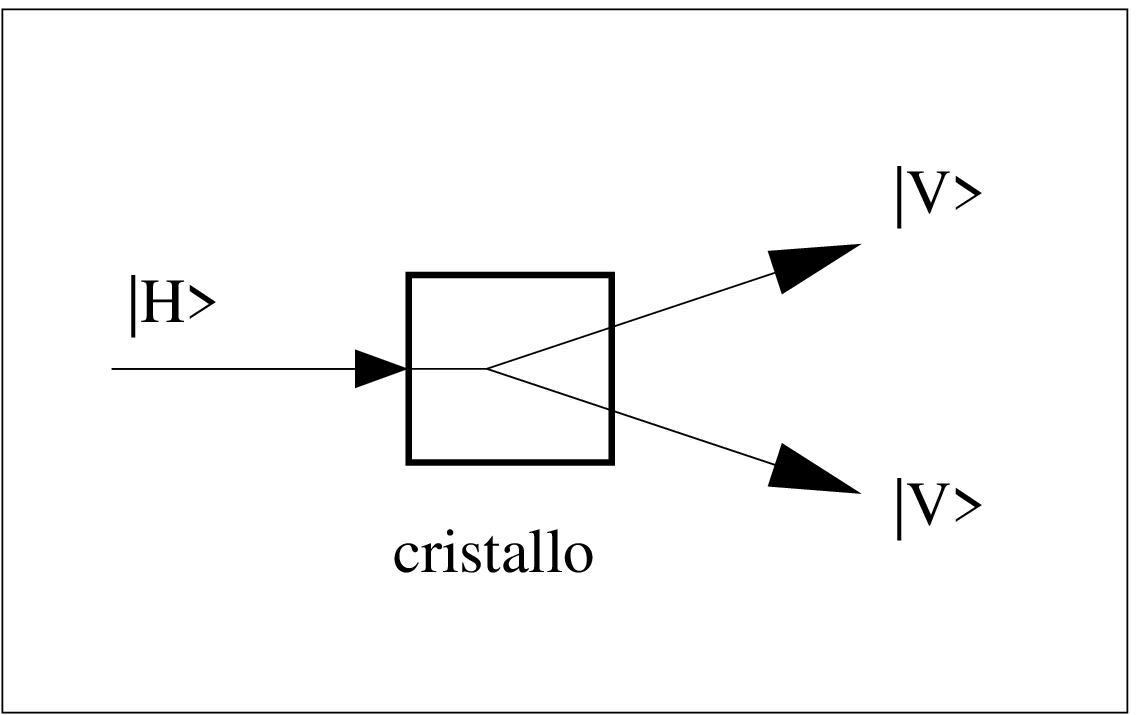}
         \begin{fig}\label{polarizzazionePDCI}
Schema indicativo della polarizzazione dei fasci
nella PDC di tipo I.
Con H si denota una polarizzazione orizzontale
e con V una verticale.
Il fascio iniettato nel cristallo è detto di pompa, 
i due emergenti $signal$ ed $idler$.
         \end{fig}
\end{minipage}
\ \hspace{5mm} \hspace{5mm} \
\begin{minipage}[b]{6cm}
   \centering
   \includegraphics[width=6cm]{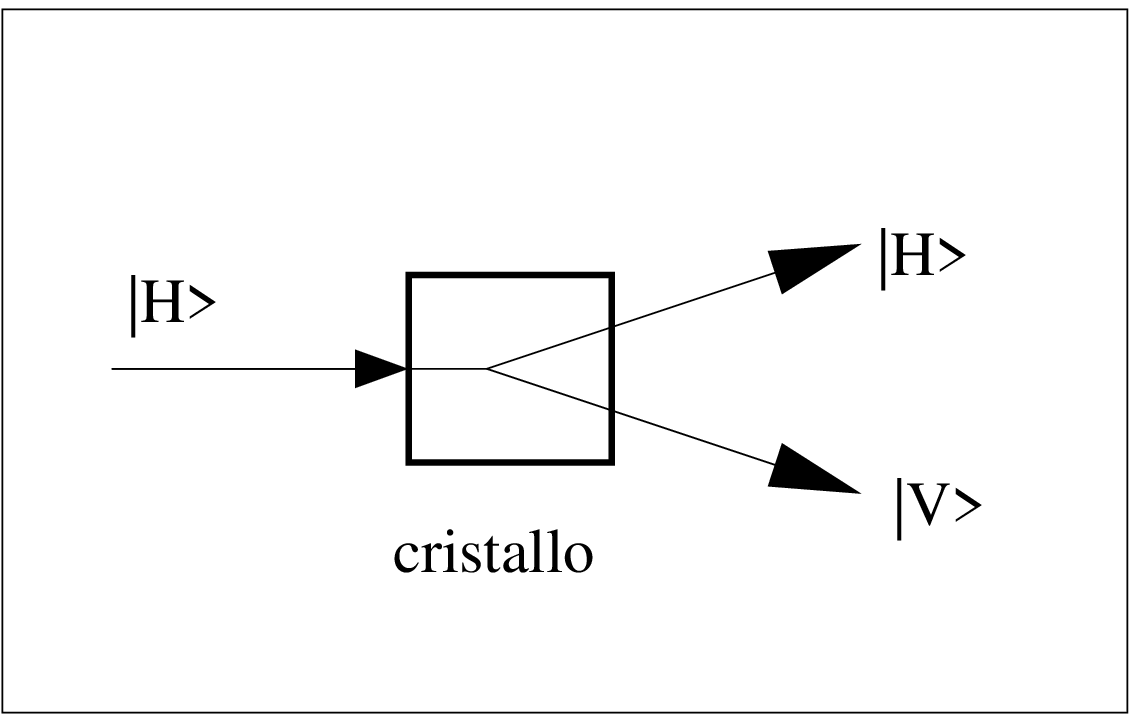}
         \begin{fig}\label{polarizzazionePDCII}
Schema indicativo della polarizzazione dei fasci
nella PDC di tipo II.
Con H si denota una polarizzazione orizzontale
e con V una verticale.
Il fascio iniettato nel cristallo è detto di pompa, 
i due emergenti $signal$ ed $idler$.
         \end{fig}
\end{minipage}
\end{figure}

Per ragioni storiche   
$\omega_1$
e
$\omega_2$
sono rispettivamente detti 
{\enfatizza idler} e {\enfatizza signal}.\\
Nella PDC di tipo I entrambi
i fotoni emessi dal cristallo hanno
polarizzazione ortogonale a quella del
fascio di pompa~(fig.~\ref{polarizzazionePDCI}).\\
Nella PDC di tipo II 
{\parolestrane signal} ed 
{\parolestrane idler}
hanno polarizzazione ortogonale fra 
loro, quindi, uno solo ha
la stessa polarizzazione del
fascio di pompa~(fig.~\ref{polarizzazionePDCII}).\\
Nel seguito ci occuperemo principalmente della PDC
di tipo~I che è stata utilizzata nel nostro esperimento.

\clearpage

\begin{figure}[h]
\begin{center}
   \includegraphics[width=14cm]{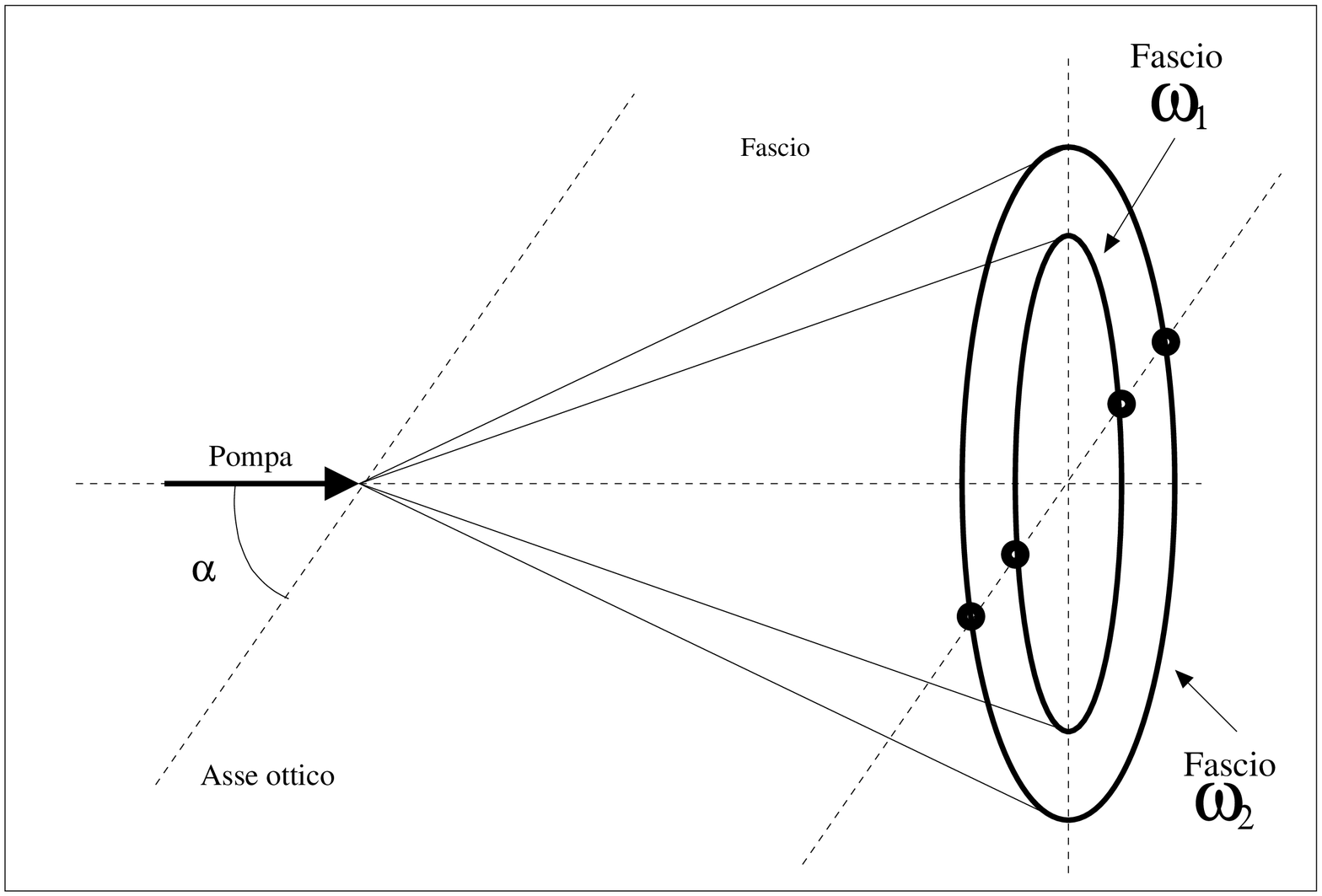}
         \begin{fig}\label{verticecono}
Geometria dell'emissione nel caso di PDC di tipo~I,
in cui si mette in rilievo la presenza di un 
angolo~$\alpha$ formato tra l'asse ottico del
cristallo e la direzione del fascio di pompa.
Le coppie di fotoni correlati 
che soddisfano le relazioni~\ref{vincolo1}-\ref{pm}
si trovano su uno stesso diametro.
         \end{fig}
\end{center}
\end{figure}

Il secondo vincolo (l'eq.~\ref{pm}) prende normalmente il nome,
in letteratura, 
di condizione di {\enfatizza phase-matching}. \\
E' possibile soddisfare tale richiesta in un mezzo 
dielettrico anisotropo attraverso 
i diversi valori di indice di rifrazione ordinario e straordinario. \\
Nel phase-matching di tipo I, i due fotoni
$\omega_1$
e
$\omega_2$
si propagano entrambi 
come onda ordinaria, quindi con polarizzazione ortogonale ad 
$\omega_3$.\\
E' evidente che le due equazioni precedenti 
(l'eq.~\ref{omega} e l'eq.~\ref{pm}) 
non determinano univocamente 
$\omega_1$
e
$\omega_2$ ; 
l'emissione di PDC è a banda larga,
cioè va dalla lunghezza d'onda di pompa all'estremo infrarosso, 
limitata solo dalla trasparenza del dielettrico usato. \\
Un'altra proprietà fondamentale 
è la simultaneità dell'emissione 
dei due i fotoni  
$\omega_1$
e
$\omega_2$,
tale caratteristica è stata verificata sperimentalmente 
a livello dei\\ femtosecondi.\\
I quanti con la stessa frequenza si trovano su
coni, il cui vertice 
è un punto all'interno del cristallo
individuato dall'intersezione
dell'asse ottico del cristallo utilizzato
con la direzione del fascio di pompa,
come illustrato in figura~\ref{verticecono}.\\
Le coppie di fotoni correlati che soddisfano le relazioni~\ref{vincolo1}-\ref{pm}
si trovano su uno stesso diametro~(fig.~\ref{verticecono}),
all'intersezione coi coni relativi alle loro rispettive frequenze.

\clearpage

La funzione d'onda che descrive lo stato che si origina per parametric down-conversion 
in una trattazione multimodale è la seguente:

\begin{eqnarray}
|\Psi(t)\rangle =\frac{1}{L^3}\frac{1}{i \hbar}%
V_{l} \sum_{[\vec K' S']_1}%
\sum_{[\vec K'' S'']_2}%
\chi_{ijl}^{(2)}(\omega_1,\omega_2,\omega_3)%
(\vec{\epsilon}^{~*}_{\vec K' S'})_1(\vec{\epsilon}^{~*}_{\vec K' S'})_2 \times \nonumber\\%
\nonumber\\%
\times \prod^3_{m=1}%
\left[ \frac{\sin \frac{1}{2} (\vec K_3 - \vec K' -\vec K'')_m l_m }%
{\frac{1}{2}  (\vec K_3 - \vec K' -\vec K'')_m } \right] %
e^{i \frac{t}{2} (\omega'+\omega''-\omega_0)} \times \nonumber\\%
\nonumber\\%
\times \frac{\sin \frac{1}{2} (\omega'+\omega''-\omega_0) t}{\frac{1}{2} (\omega'+\omega''-\omega_0)}%
|\vec K', S' \rangle _1 |\vec K'',S''\rangle _2\nonumber\\
\end{eqnarray}

dove $L^3$ è il volume di quantizzazione,
$S$ sono gli stati di polarizzazione,
$\vec K$ è il numero d'onda,
$l_m$ (con m=x, y, z) indica le dimensioni del cristallo,
$V_{l}$ è l'ampiezza del campo di pompa,
$\chi_{ijl}^{(2)}$ è la suscettività e gli
$\vec \epsilon$  sono i vettori di polarizzazione.\\
La caratteristica più importante dell'espressione precedente è che, 
a causa delle sommatorie, 
lo stato 
$|\Psi\rangle$ non è fattorizzabile nel prodotto delle funzioni d'onda dei due singoli fotoni 
$|\omega \rangle _1$
e 
$|\omega \rangle _2$,
ovvero rappresenta uno stato entangled.

\section{Descrizione teorica della fluorescenza parametrica}

Per una trattazione più dettagliata del fenomeno della PDC si consideri
l'Hamiltoniana del campo elettromagnetico, la quale
in un ambito classico ha la seguente espressione:

\begin{equation}\label{hamiltonianaI}
H=\int\frac{1}{2\mu_0}\vec{B}^2(r,t)d^3r+\int d^3r\int^{D(r,t)}_0 \vec{E}(r,t) \cdot d\vec{D}(r,t)
\end{equation}

dove $\vec B$ è il vettore induzione magnetica,
$\vec E$ il vettore campo elettrico e 
$\vec D$ è il vettore induzione dielettrico.\\ 
Nel caso in cui la radiazione si propaghi in uno spazio vuoto 
o in un mezzo isotropo il secondo  integrando
assume la forma semplificata:

\begin{equation}
\frac{1}{2}\vec{E}(r,t) \cdot \vec{D}(r,t)
\end{equation}

In generale la relazione che lega $\vec{D}$ ed $\vec{E}$ non è banale,
in un mezzo non lineare $\vec{D}(r,t)$ è espresso come:

\begin{equation}
\vec{D}(r,t)= \epsilon_0 \vec{E}(r,t)+\vec{P}(r,t)
\end{equation}

Dove $\vec{P}(r,t)$ è il vettore di polarizzazione che può essere sviluppato in serie di potenze 
in $\vec{E}$:

\begin{equation}
P_i= \chi_{ij}^{(1)}E_j+\chi_{ijk}^{(2)}E_jE_k+\chi_{ijkl}^{(3)}E_jE_kE_l+...
\end{equation}

In cui $\chi^{(n)}$ è il tensore suscettività elettrica di rango $n+1$.

In ambito quantistico, 
l'Hamiltoniana~\ref{hamiltonianaI}
può essere riscritta come
una somma di tre modi di oscillazione libera del campo elettromagnetico (quello
del campo di pompa e quelli dei due campi 
{\parolestrane signal} e 
{\parolestrane idler}) e di un termine
d'interazione che descrive l'annichilazione del fotone di pompa e la
creazione dei due fotoni 
{\parolestrane signal} e 
{\parolestrane idler}, pi\`{u} il suo Hermitiano coniugato
per il processo inverso: 

\begin{equation}
\hat{H}=\sum _{i=1}^{3}\hbar\omega _i%
\left(\hat{n}_i+\frac{1}{2}\right)+%
\hbar g\left[\hat{a}_1^{\dagger}\hat{a}_{2}^{\dagger}\hat{a}_3+%
h.c.\right]
\label{hamiltoniana}
\end{equation}

dove $\hat{a}_1$, $\hat{a}_2$,  $\hat{a}_3$
sono gli operatori di campo che creano,
rispettivamente, un fotone $signal$,
un fotone $idler$ ed un fotone di pompa;
il coefficiente d'accoppiamento $g$ tiene conto delle propriet\`{a} non lineari del
cristallo ed è quindi legato alla 
suscettivit\`{a} elettrica non lineare del mezzo.
Si verifica che:

\begin{equation}
\left[\hat{n}_1+\hat{n}_2+2\hat{n}_3,\hat{H}\right]=0  
\label{commutatore}
\end{equation}

ovvero 
$\hat{n}_1+\hat{n}_2+2\hat{n}_3$
 \`{e} una costante del moto, il che
esprime il `decadimento' del fotone di pompa in un fotone 
{\parolestrane signal} e un
fotone {\parolestrane idler}.

E' possibile ottenere una soluzione analitica per l'evoluzione temporale
degli operatori $\hat{a}_i(t)$ con $i=1,2,3$, 
 introducendo una semplificazione di $\hat{H}$: supposto
che il campo incidente sia tanto intenso  da poter essere descritto
classicamente come un campo di ampiezza complessa 
$a_0=v_0e^{-i\omega_0t}$, si pu\`{o} eliminare uno dei modi quantizzati dall'Hamiltoniana,
ottenendo: 

\begin{equation}
\hat{H}=\sum _{i=1}^{2}\hbar\omega _i%
\left(\hat{n}_i+\frac{1}{2}\right)%
+\hbar g\left[\hat{a}_1^{\dagger}\hat{a}_2^{\dagger}v_0e^{-i\omega_0t}%
+h.c.\right]
\end{equation}

Valida sotto la condizione:

\begin{equation}
\left\langle \hat{n}_{1}(t)\right\rangle ,\left\langle \hat{n}_{2}(t)\right\rangle<<\left\langle n_{0}\right\rangle =\left\langle a_{0}^{\dagger}a_{0}\right\rangle =\left| v_{0}\right| ^{2}
\end{equation}

In condizioni di
campo di pompa costante
l'approssimazione fatta
è garantita.
In questo caso si ottiene che $\hat{n}_1(t)-\hat{n}_2(t)$ \`{e} una costante del moto:

\begin{equation}
\left[\hat{n}_1(t)-\hat{n}_2(t),\hat{H}\right]=0
\end{equation}

Il che esprime il fatto che i due fotoni 
{\parolestrane signal} e 
{\parolestrane idler} sono sempre creati
insieme.\\

Le equazioni del moto, 
in formulazione alla Heisenberg, 
per gli operatori $\hat{a}_1$ e $\hat{a}_2$
sono: 

\begin{equation}
\frac{d\hat{a}_1(t)}{dt}=\frac{1}{i\hbar}%
\left[\hat{a}_1(t),\hat{H}\right]%
=-i\omega_1\hat{a}_1(t)-ig\hat{a}_2^{\dagger}(t)v_0e^{-i\omega _0t}
\end{equation}

\begin{equation}
\frac{d\hat{a}_2(t)}{dt}=\frac{1}{i\hbar}%
\left[\hat{a}_{2}(t),\hat{H}\right]%
=-i\omega_2a_2(t)-iga_1^{\dagger}(t)v_0e^{-i\omega _0t}
\end{equation}

Moltiplicando $a_{l}$ per $e^{i\omega _{l}t}$ 
$(l=1,2)$ si
ottengono operatori $\hat{A}_{l}(t)=a_{l}(t)e^{i\omega _{l}t}$
che evolvono secondo le equazioni:

\begin{equation}
\frac{d\hat{A}_1(t)}{dt}=-igv_0\hat{A}_2^{\dagger }(t)e^{i(\omega _1+\omega_2-\omega _0)t}
\end{equation}

\begin{equation}
\frac{d\hat{A}_2(t)}{dt}=-igv_0\hat{A}_1^{\dagger }(t)e^{i(\omega _1+\omega_2-\omega _0)t}
\end{equation}

Nel caso particolare $\omega _1+\omega _2=\omega _0$
si ottengono due equazioni accoppiate: 

\begin{equation}
\frac{d\hat{A}_1(t)}{dt}=-igv_0\hat{A}_2^{\dagger}(t)
\end{equation}

\begin{equation}
\frac{d\hat{A}_2(t)}{dt}=-igv_0\hat{A}_1^{\dagger}(t)
\end{equation}

per disaccoppiare le quali occorre derivare una seconda volta:

\begin{equation}
\frac{d^2\hat{A}_1}{dt^2}=g^2|v_0|^2\hat{A}_1^{\dagger}(t)
\end{equation}

\begin{equation}
\frac{d^2\hat{A}_2}{dt^2}=g^2|v_0|^2\hat{A}_2^{\dagger}(t)
\end{equation}

Imposte le opportune 
condizioni inziali,
le soluzioni sono:

\begin{equation}
\hat{A}_1(t)=\hat{A}_1(0)\cosh(g|v_0|t)-ie^{i\theta}\hat{A}_2^{\dagger}(0)\sinh(g|v_0|t) 
\end{equation}

\begin{equation}
\hat{A}_2(t)=\hat{A}_2(0)\cosh(g|v_0|t)-ie^{i\theta}\hat{A}_1^{\dagger}(0)\sinh(g|v_0|t) 
\end{equation}

dove  $v_0=|v_0|e^{i\theta}$.

Nota l'evoluzione degli operatori, \`{e} possibile calcolare i momenti
statistici di ordine $r$ per gli operatori numero $\hat{n}_1(t)$ e $\hat{n}_2(t)$.
Assumendo
come stato iniziale dei campi 
{\parolestrane signal} e 
{\parolestrane idler} il prodotto degli stati di
vuoto $|vac\rangle_1|vac\rangle_2$ e
indicando con~`:~:' il prodotto normale dei campi, si ottiene: 

\begin{equation}
\left\langle \hat{n}_{j}^{(r)}(t)\right\rangle =\left\langle :\hat{n}_j^r(t):\right\rangle=_{1,2}\langle vac|\hat{A}_j^{\dagger r}(t)\hat{A}_j^r(t)|vac\rangle _{1,2}
\end{equation}

dove $j=1,2$.\\
Da cui, dopo alcuni passaggi:

\begin{equation}
\left\langle \hat{n}_j^{(r)}(t)\right\rangle=r!\sinh^{2r}(g|v_0|t) \label{I}\label{I}
\end{equation}

I momenti ottenuti sono identificativi della statistica
di Bose-Einstein, che descrive
fotoni emessi da sorgenti in equilibrio.\\
Si consideri un caso particolare dell'espressione
\ref{I}, cioé $r=1$:\\

\begin{equation}
\label{II}
\langle \hat{n}_1(t)\rangle 
=\sinh^2(g|v_0|t)
=\langle \hat{n}_2(t)\rangle 
\end{equation}

Il numero di fotoni prodotti per PDC varia nel tempo
ed è dato dall'equazione~\ref{II}
per $g|v_0|t\ll 1$.\\ 
Per tempi maggiori la crescita diviene
esponenziale, ma in questo caso l'approssimazione fatta (che permette solo
valori di 
$\langle \hat{n}_{j}\rangle |v_0|^2$ 
prossimi a
zero) non sarebbe più valida. Il tempo d'interazione $t$ 
\`{e} il tempo di propagazione nel mezzo non lineare, che di solito \`{e} cos%
\`{i} breve da giustificare l'assunzione  
$g|v_0|t\ll 1$.

Si pu\`{o}, quindi anche ricavare la correlazione incrociata~(dove $j=1,2$):

\begin{equation}
\langle :\hat{n}_1(t)\hat{n}_2(t):\rangle =
\langle vac|\hat{A}_1^{\dagger}(t)\hat{A}_2^{\dagger}(t)\hat{A}_2(t)\hat{A}_1(t)|vac\rangle_{1,2}=
\end{equation}

\begin{equation}
=\langle \hat{n}_j(t)\rangle+%
2\langle \hat{n}_j(t)\rangle^2 =
\end{equation}

\begin{equation}%
\label{III}
=\langle \hat{n}_j(t)\rangle+%
\langle :\hat{n}_j^2(t):\rangle 
\end{equation}

La correlazione incrociata delle fluttuazioni del numero di fotoni è data:\\

\begin{equation}
\langle :\Delta\hat{n}_1(t)\Delta\hat{n}_2(t):\rangle =
\langle :\hat{n}_1(t)\hat{n}_2(t):\rangle-
\langle \hat{n}_1(t)\rangle
\langle \hat{n}_2(t)\rangle
\end{equation}

Da questa espressione si ottiene il coefficiente 
di correlazione incrociata\\
normalizzato~$\sigma_{12}$:\\

\begin{equation}
\sigma _{12}\equiv\frac{\langle :\Delta \hat{n}_1(t)\Delta \hat{n}_2(t):\rangle}%
{\sqrt{\langle(\Delta \hat{n}_1(t))^2\rangle \langle ( \Delta \hat{n}_2(t))^2\rangle}}=1
\end{equation}

I segnali {\parolestrane signal} e 
{\parolestrane idler} sono dunque massimamente correlati,
ogni incremento nei fotoni {\parolestrane signal} 
corrisponde ad un uguale incremento nei fotoni
{\parolestrane idler}.

\chapter{Esperimento}

\section{Alcuni precedenti esperimenti con doppia fenditura e PDC}\label{esperimenticondoppiafenditura}

\begin{figure}[h]
\begin{center}
\includegraphics[width=8cm]{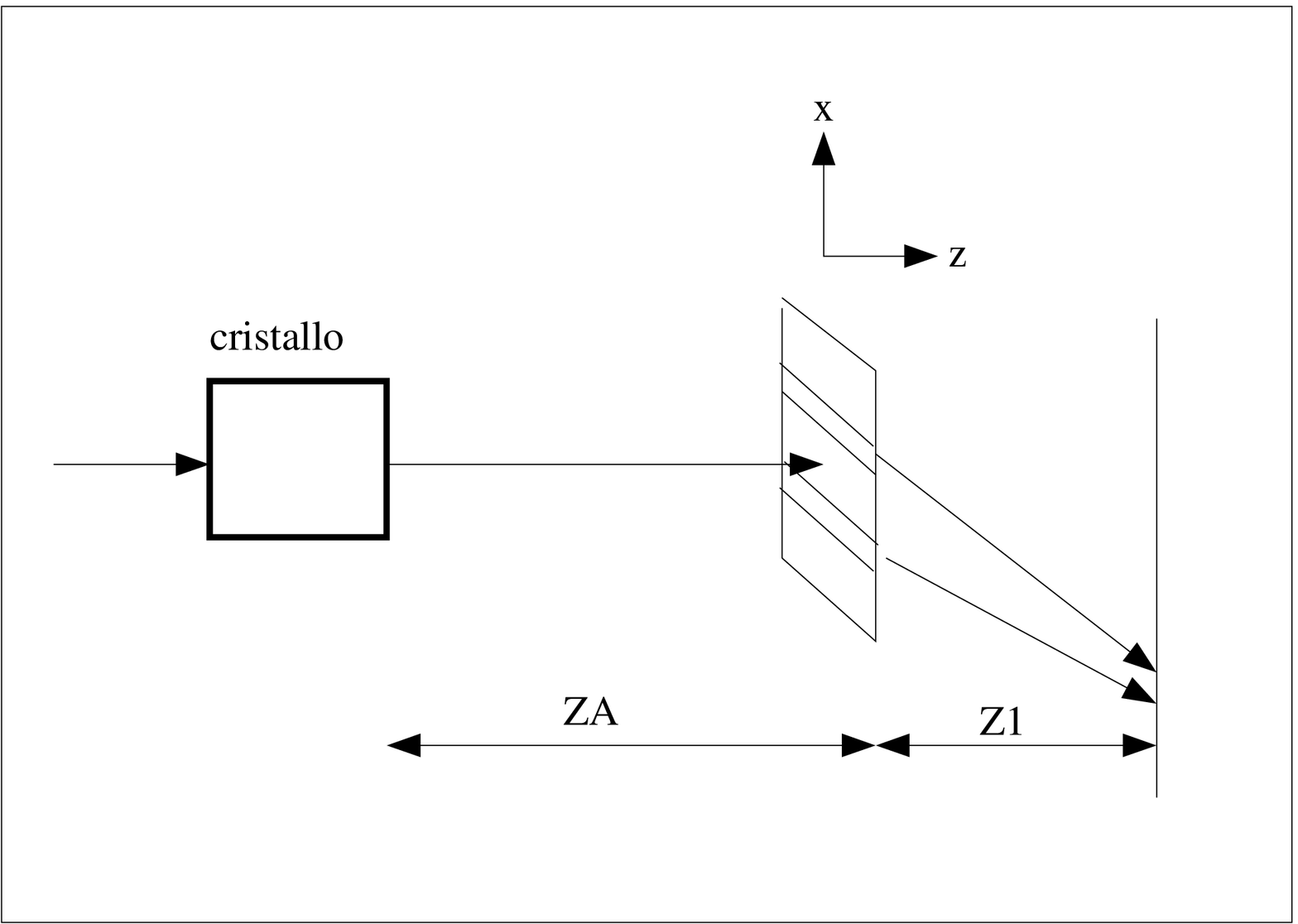}
\end{center}
    \begin{fig}\label{storta}
    \begin{center}
    Schema dell'esperimento a doppia fenditura di ref.\cite{2}.
    \end{center}
    \end{fig}
\end{figure}

Uno degli aspetti caratteristici della fisica quantistica
è rappresentato dal principio di complementarità,
in base al quale in un esperimento in cui si mette
in rilievo la natura ondulatoria di un sistema quantistico
non si possono osservare gli effetti dovuti all'aspetto 
corpuscolare e viceversa.\\
Come illustrato nell'esperimento citato nell'introduzione
l'analisi delle frange di interferenza 
prodotte dal passaggio di fotoni attraverso una 
doppia fenditura può fornire informazioni interessanti
a questo proposito.\\
A tal fine, in questi ultimi anni, sono state realizzate
due tipologie di esperimenti,
con fotoni prodotti in PDC polarizzati\footnote{
In tutti questi esperimenti i fotoni prodotti a coppie 
sono degeneri , ovvero, con la stessa lunghezza d'onda.
} 
ed una doppia fenditura:
in una prima classe di esperienze~\cite{2} e~\cite{1},
sia i fasci
{\parolestrane signal}
che
{\parolestrane idler}
vengono inviati
verso la doppia fenditura,  
negli altri la doppia fenditura è inserita su un solo ramo,
ad esempio corrispondente al 
{\parolestrane signal}~\cite{3}, \cite{4}, \cite{5}.\\ 
Alla prima classe di esperienze appartiene
l'esperimento realizzato da
Fonseca ed altri~\cite{2}.
Come illustrato in 
fig.~\ref{storta}
entrambe le emissioni prodotte in un cristallo
di Beta Borato di Bario
(in cui si inietta un fascio di pompa
di lunghezza d'onda 351.1~$nm$, generato da un
laser ad Argon con una potenza di
400~$mW$, al fine di produrre fluorescenza 
parametrica di tipo~II)
vengono inviati verso la stessa fenditura, 
il numero di pacchetti a due fotoni
che vi passano attraverso raggiungendo il piano di detezione
è proporzionale al quarto ordine della funzione di
correlazione calcolata  nel punto $x$:

\begin{equation}\label{quartoordine}
N_c(x)\varpropto \langle%
\hat{E}^-_i(x)%
\hat{E}^-_s(x)%
\hat{E}^+_i(x)%
\hat{E}^+_s(x)%
\rangle
\end{equation}

Dove $\hat{E}^+_i(x)$
e $\hat{E}^+_s(x)$
sono gli operatori di campo elettrico trasmessi del 
{\parolestrane signal} e 
dell'{\parolestrane idler}.
La presenza di quattro operatori nella 
formula~\ref{quartoordine} denota uno studio 
del fenomeno al quarto ordine.
E' necessario uno studio dell'interferenza
al quarto ordine poiché al secondo ordine non si evidenzia
la sovrapposizione
dell'interferenza prodotta dai due singoli quanti
con quella del pacchetto nel suo complesso,
scopo dell'esperienza che si sta considerando.\\
In ref.~\cite{2} tale equazione è stata calcolata
usando la funzione d'onda multimodale di due fotoni,
in approssimazione monocromatica.
Dalla~\ref{quartoordine} segue che
il numero di coppie di fotoni entangled,
in funzione della posizione $x$ è dato da:

\begin{eqnarray}\label{coincidenze2}
N_c(x) \varpropto A(x)+%
4B_1(x)B_2(x)\cos\left(\frac{kd^2}{z_A}+\frac{kx2d}{z_1}\right)+\nonumber\\
+4B_2(x)B_4(x)\cos\left(\frac{kd^2}{z_A}+\frac{kx2d}{z_1}\right)+\nonumber\\
+4B_1(x)B_4(x)\cos\left(\frac{2kx2d}{z_1}\right)
\end{eqnarray}

Dove $2d$ è la distanza tra le due fenditure,
$2a$ è la larghezza di ogni apertura,
$z_1$ la distanza della doppia fenditura dai fotorivelatori
e $z_A$ è la distanza longitudinale dal cristallo
(fig.~\ref{storta}).
I termini $B_1(x)$, $B_2(x)$, $B_4(x)$ e $A(x)$
sono coefficienti che dipendono dalla posizione $x$,
per una loro definizione~ref.\cite{2}.\\ 
L'equazione~\ref{coincidenze2} presenta più termini:
il secondo ed il terzo 
derivano dall'interferenza di fotoni individuali
({\parolestrane signal} e {\parolestrane idler}) 
ed hanno periodo di oscillazione 
corrispondente a $\lambda_0$=702.2~nm;
il quarto 
descrive una figura di interferenza 
generata da un fascio di pompa con
lunghezza d'onda 
$\frac{\lambda_0}{2}$.\\
L'apparato strumentale utilizzato consente anche di realizzare
una situazione intermedia,
ove l'energia dello stato quantistico è ben
definita, ma non è possibile misurare la 
lunghezza d'onda di de~Broglie.
I dati ottenuti sono in accordo
con la previsione teorica (eq.~\ref{coincidenze2}).\\
Risultato fondamentale di tale studio
è, dunque, che modificando il profilo trasverso
del campo bifotonico alla posizione
della doppia fenditura  si transisce da una situazione ove
si osserva l'interferenza del singolo fotone
(e in cui si misura una lunghezza d'onda $\lambda_0$)
ad una ove si considera l'interferenza del campo bifotonico
nel suo complesso (con `lunghezza d'onda' $\lambda_0/2$).

Sempre nel contesto di questa prima classe di esperimenti,
un lavoro teorico è stato condotto
da Abouraddy ed altri~\cite{1}.
Esso consiste in una simulazione
del fenomeno dell'interferenza
al variare dello spessore del cristallo
(si eseguono simulazioni per 0,1~mm; 1~mm e 10~mm)
e, contemporaneamente, dell'orientazione del cristallo,
variando l'angolo formato tra l'asse ottico 
e la direzione del fascio di pompa
($36.30^0$; $36.50^0$ e $36.44^0$).
In questo caso si ipotizza 
di utilizzare un cristallo di BBO
in cui il fascio di pompa è prodotto da un laser
di He-Cd con una lunghezza d'onda 325 nm; i fotoni 
{\parolestrane signal} ed
{\parolestrane idler} 
sono indirizzati su distinte fenditure.\\
L'intensità dell'interferenza prodotta da fotoni
entangled è descritta dalla funzione di coerenza
al quarto ordine, che si verifica avere una forma simile a quella al secondo
ordine 
di un sistema che utilizza sorgenti classiche parzialmente coerenti.
Da tali simulazioni emerge che la realizzazione
di un esperimento che utilizzi 
questo tipo di configurazione
presenta più gradi di libertà di un 
analogo classico
(con sorgenti non parametriche).
Si è osservato, inoltre, che modificando lo spessore
del cristallo, la direzione dell'asse ottico
o l'ampiezza di banda del sistema,
lasciando fissa la distanza tra le fenditure,
si può traslare la figura di interferenza.\\
Una seconda classe di esperimenti è indirizzata
allo studio del caso in cui
solo uno dei fasci attraversa la doppia fenditura,
con lo scopo di studiare le correlazioni
quantistiche tra i due rami.\\
Nell'esperimento discusso in~\cite{3}
solo il fascio {\parolestrane signal} 
è inviato verso la doppia fenditura,
mentre sul percorso dell'altro ({\parolestrane idler}) 
è stata inserita un'iride
di diametro variabile al fine di valutare
la variazione della visibilità
delle frange di interferenza
al quarto ordine (coincidenze).
Variando il diametro dell'iride
sul fascio {\parolestrane idler} si selezionano
i vettori d'onda $\vec k_i$
del fascio {\parolestrane idler} stesso.
Grazie alla forte correlazione 
tra i due fotoni gemelli
questo implica anche una selezione
dei vettori d'onda sul ramo 
{\parolestrane signal}
qualora si effettui una misura di 
coincidenze.
Riducendo il diametro dell'iride sul ramo {\parolestrane idler}
si osserva, quindi, l'attenuazione delle
frange di interferenza coerentemente col
principio di complementarità.\\
Lo studio di ref.~\cite{3} è basato su una legge
empirica; al fine di fornire un supporto teorico più rigoroso
un'altro gruppo di ricerca~\cite{4} ha sviluppato
il formalismo necessario a definire il coefficiente di visibilità
partendo dalla funzione di correlazione.
Tale coefficiente valutato al quarto ordine per una sorgente
coerente quantistica ha la stessa forma di
quello al secondo ordine per una sorgente classica
quasi coerente, come già accennato in~\cite{1}.\\
Infine, lo studio discusso in ref.~\cite{5} è analogo ai due precedenti,
ma a differenza di questi non è stata posta l'iride
sul fascio $idler$ ed oltre all'analisi della figura
di interferenza si è considerata quella prodotta
dalla diffrazione.
I risultati ottenuti sperimentalmente sono
compatibili con le previsioni quantistiche.

\section{L'esperimento allo IENGF}

\begin{figure}[h]
\begin{center}
\includegraphics[width=4cm]{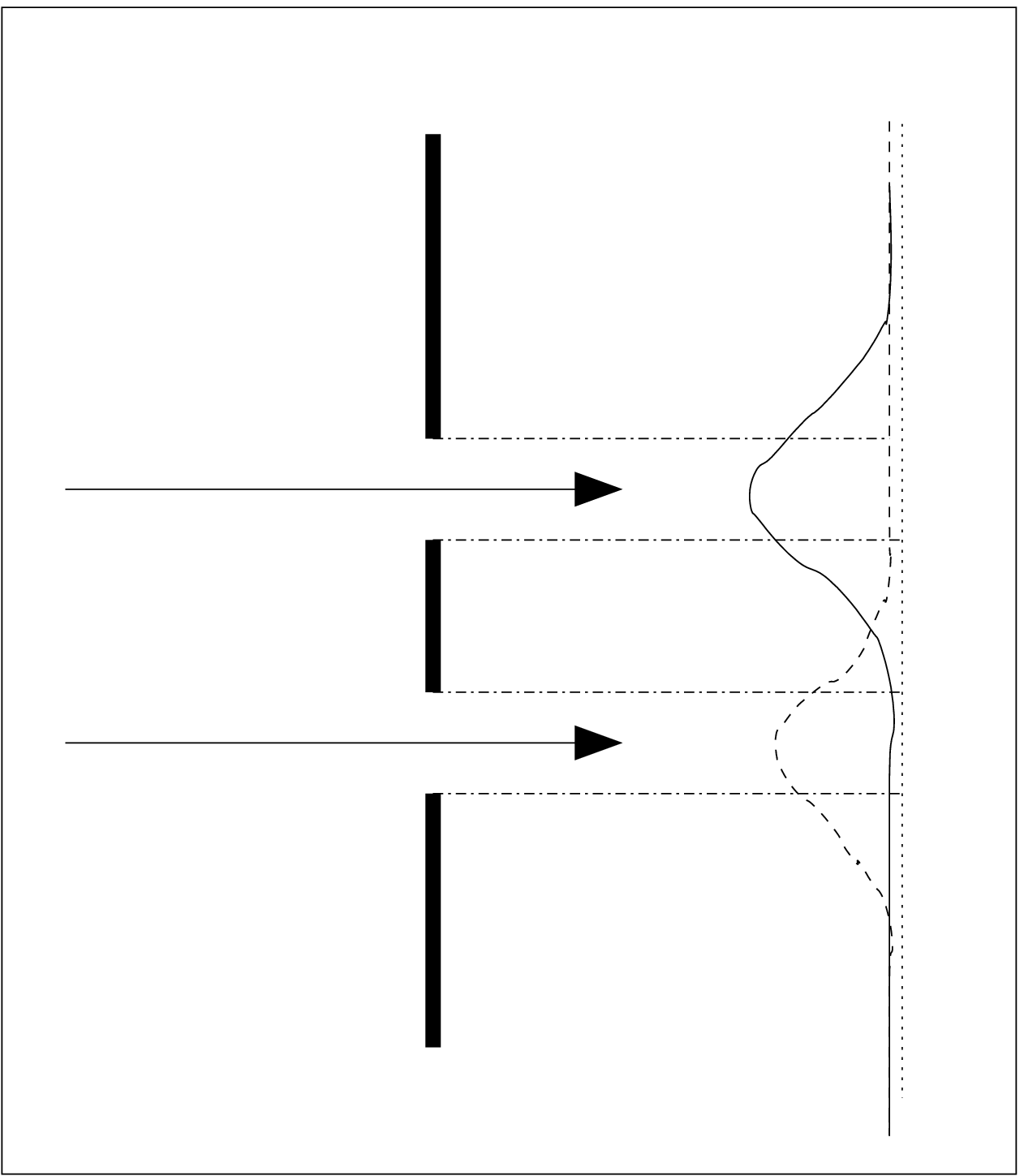}
\end{center}
     \begin{fig}\label{doppiafenditura}
Esempio di
funzioni d'onda di due singoli fotoni che attraversano la 
doppia fenditura ognuno per una determinata apertura.
     \end{fig}
\end{figure}

Come già illustrato precedentemente,
gli esperimenti basati sulla doppia fenditura consentono 
uno studio approfondito del principio di
complementarità e delle sue conseguenze.
In questo contesto risultano particolarmente significativi
gli studi realizzati mediante campi 
{\parolestrane bifotonici}, cioè, non mere coppie di
singoli quanti, ma particelle 
{\parolestrane entangled}, 
quindi, strettamente correlate in direzione e impulso.\\
Nell'esperimento analizzato in questa tesi
due fotoni indistinguibili
(perché aventi entrambi lunghezza d'onda 702~$nm$,
e stessa polarizzazione in quanto prodotti per 
PDC di tipo~I) sono vincolati a passare attraverso 
una doppia fenditura,
ognuno per una determinata apertura
(a differenza di quanto realizzato da~\cite{2}).\\
Il pacchetto d'onda che descrive il singolo quanto è
estremamente stretto, quindi, è possibile ottenere una 
configurazione sperimentale
in cui la $\psi$  associata alla particella che passa 
attraverso una fenditura sia nulla in corrispondenza
della seconda~(fig.~\ref{doppiafenditura}).
Tale caratteristica consente 
uno studio innovativo della figura di interferenza,
in quanto, il contributo all'interferenza
dovuto ai singoli quanti che costituiscono
il pacchetto è soppresso,
ed è possibile valutare l'interferenza
al quarto ordine dovuta solo
al pacchetto bifotonico nel suo complesso.
I risultati del nostro lavoro sono riportati
nel paragrafo~\ref{acquisizioneedanalisidati}.\\
La proprietà di entanglement, che caratterizza i fotoni
prodotti per fluorescenza parametrica,
e la correlazione temporale che esiste tra il 
{\parolestrane signal} e l'{\parolestrane idler}, 
permettono di ottenere la 
presenza dei due fotoni allo stesso
istante, rispettivamente,
nelle due fenditure\footnote{
Nell'ipotesi di un corretto posizionamento
della doppia fenditura, operazione 
non semplice da realizzare, che verrà 
descritta nel seguito.
}.
Tale proprietà risulta fondamentale nella realizzazione 
del test sperimentale sulla teoria di de~Broglie-Bhom,
discussa nel capitolo~\ref{dBB}.

\section{Descrizione dell'esperimento.}

Si utilizza come sorgente di pompa un laser ad Argon
che emette a 351~$nm$, con una potenza
massima di 0.8~W.
Il tubo a bassa pressione (0.001~atm) contenente il gas
è posizionato all'interno di una cavità a due
specchi.\\
Il materiale attivo, il gas nobile Argon,
è ionizzato da un impulso iniziale ad alta tensione.
La scarica elettrica fornisce energia
agli ioni di Argon i cui elettroni transiscono
ad un livello energetico metastabile più alto
per poi diseccitarsi e decadere su un livello 
stabile più basso, generando l'emissione di 
radiazione, che diventa luce laser quando
il sistema atomico è posto in un risuonatore ottico.
Il laser può oscillare su diverse lunghezze d'onda,
ognuna prodotta dalla transizione di elettroni
tra diverse coppie di livelli energetici.
Le lunghezze d'onda superiori ai 400~$nm$
sono prodotte da ioni di $Ar^+$;
quelle inferiori ai 400~$nm$ da ioni di $Ar^{2+}$.
La lunghezza d'onda utilizzata ai fini 
dell'esperimento è $\lambda=351.1$~$nm$,
selezionata dal risuonatore.
La larghezza dell'emissione di questo laser è
di circa 6~$GHz$.
Il fascio emesso è polarizzato verticalmente.\\
Sul percorso ottico del fascio viene, collocata una lente
biconvessa con lunghezza focale 2.1~m allo scopo di diminuire il diametro dello
spot del laser in corrispondenza del cristallo non-lineare.\\
Per esigenze pratiche il laser è posizionato su un banco ottico
diverso da quello di lavoro, è quindi necessario
usare un sistema di specchi per dirigere il fascio, 
su un cristallo di~$LiO_3$, con una inclinazione di $51^0$
rispetto l'asse ottico.\\
Come illustrato in dettaglio nel capitolo~\ref{entanglement},
iniettando un fascio laser di pompa all'interno di un opportuno
mezzo non lineare (il cristallo di~$LiO_3$)
si generano, per PDC di tipo~I,
fotoni fortemente correlati in frequenza, impulso e 
polarizzazione~\footnote{
Come spiegato nel capitolo~\ref{entanglement}
ciò è possibile solo con la PDC di tipo~I.
}
che vengono emessi in direzioni diverse
in funzione della loro lunghezza d'onda.\\
A causa della necessità di avere un asse di simmetria ben definito
e fotoni indistinguibili si scelgono fasci 
{\parolestrane signal} e
{\parolestrane idler}
con identica frequenza;
essendo il fascio di pompa a 351~$nm$,
la legge di conservazione dell'energia
impone l'uso,
per questo esperimento,
di fotoni a 702~$nm$.\\
Dopo il cristallo si introduce un filtro~UV per eliminare
il fascio di pompa residuo che creerebbe rumore 
di fondo durante l'acquisizione dati.\\
Al fine di far convergere la fluorescenza parametrica 
uscente dal cristallo 
nel punto in cui è posizionata la doppia fenditura,
si utilizza un condensatore ottico costituito da una coppia 
di lenti piano convesse coassiali.
Il sistema ottico così costituito ha due fuochi simmetrici
distanti 125~$mm$.
Sull'asse delle lenti si è, inoltre, praticato un foro
che permette il passaggio del fascio~UV 
evitando emissioni di fluorescenza
dalle lenti stesse.\\
Le due fenditure sono state realizzate
mediante  deposizione di niobio
con processo litografico su un vetrino;
le dimensioni scelte sono:
10~$\mu m$ di larghezza,
1~$mm$ di altezza ad una distanza reciproca di 100~$\mu m$.
Ognuno dei due fotoni della coppia entangled deve
passare attraverso una specifica fenditura le cui dimensioni devono
essere tali da poter osservare gli effetti
di diffrazione.
La distanza di 100~$\mu m$ è stata scelta per permettere 
di ottenere una figura di interferenza opportuna.\\
Infine, per effettuare un'ulteriore selezione spettrale  sui fotoni
si utilizzano filtri interferenziali a 702 $nm$,
con FWHM\footnote{Full width at half height, ovvero,
larghezza a mezza altezza.
} 
di 4~$nm$, i quali, per esigenze sperimentali,
sono stati montati su supporti solidali con i rilevatori.

La figura di interferenza e di diffrazione che i fotoni generano
è stata studiata mediante l'uso di un apparato strumentale,
il quale permette 
di misurare sia i conteggi di singolo fotone sui due
canali, sia le coincidenze tra questi, ossia le coppie di
fotoni che raggiungono i due rilevatori simultaneamente.\\

\clearpage

\section{Apparato di rilevazione}\label{apparatodirilevazione}

\begin{figure}[h]
\begin{center}
\includegraphics[width=8cm]{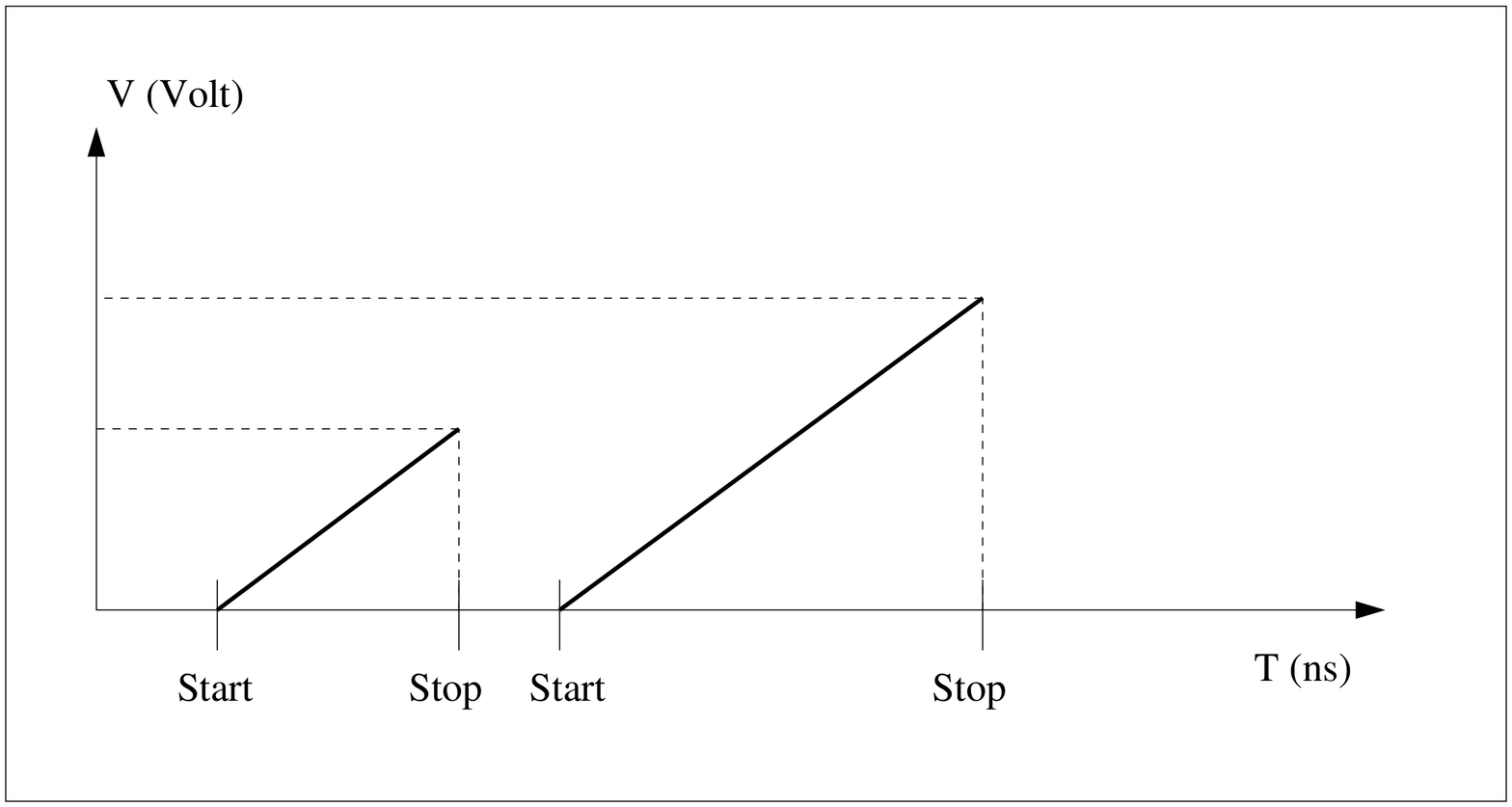}
      \begin{fig}\label{rampa}
Rampa di potenziale generata dall'elettronica interna
al TAC.
      \end{fig}
\end{center}
\end{figure}

Al fine di studiare la figura di interferenza 
e di diffrazione prodotta
dai fotoni che passano attraverso la doppia fenditura
si è utilizzata una coppia di
rilevatori di singolo fotone.\\
Essi si basano su dei fotodiodi al silicio
con superficie sensibile molto piccola,
del diametro di 140~$\mu$m.\\
Per aumentare l'area di raccolta si è posto
davanti alla superficie sensibile di ciascun
rilevatore una lente del diametro di 6~$mm$
con rivestimento anti-riflesso % sostituire coatde
per 702~$nm$.\\
Il rilevatore utilizzato per questo esperimento
ha un intervallo di detezione per fotoni
con lunghezza d'onda compresa tra 400~$nm$
e 1060~$nm$.
Quando questi incidono sulla superficie sensibile,
il circuito interno genera un impulso di tipo
TTL (Transistor-Transistor-Logic)
di ampiezza 2.5~volt (minimo),
su un carico resistivo di 50~$\Omega$ di
durata di 5~ns.\\
Alcuni segnali possono essere prodotti
anche in assenza di cattura di un quanto
e sono dovuti solo al rumore dell'elettronica interna,
i conteggi che ne risultano sono detti di buio
e sono stati stimati nell'ordine dei
700~conteggi al secondo per un fotorivelatore
e 35~per l'altro.\\
Entrambi i dispositivi possono contare 
fino a un massimo di 10~milioni di fotoni al secondo.
Il tempo morto, cioè l'intervallo che intercorre tra 
l'emissione dell'impulso e 
l'attivazione per la rilevazione del fotone successivo,
è di 50~ns.\\ 
I segnali dei due rilevatori vengono
inviati ad un  TAC-SCA\footnote{Il
dispositivo utilizzato unisce un 
Time Amplitude Converter~(TAC) 
ed un Single Channel Analizer~(SCA).},
mediante il quale è possibile individuare
le coincidenze, ovvero quei conteggi
prodotti in seguito alla simultanea
cattura di un quanto da parte di 
entrambi i fotorivelatori.\\
Il Time Amplitude Converter invia in uscita
un impulso di tensione proporzionale all'intervallo di tempo
trascorso tra l'arrivo dell'impulso di start
(dato dal rilevatore che ha un numero di conteggi
al secondo più basso) e l'impulso di stop proveniente 
dal secondo, opportunamente ritardato
mediante una linea di ritardo.\\
Quando arriva il segnale di start,
l'elettronica interna al TAC 
fa partire una rampa di potenziale~(fig.~\ref{rampa})
successivamente chiusa dall'impulso di stop,
se questo è prodotto con un ritardo inferiore ai 20~$ns$
impostati, in caso contrario si ha comunque l'interruzione.\\

\begin{figure}[h]
\begin{center}
\includegraphics[width=8cm]{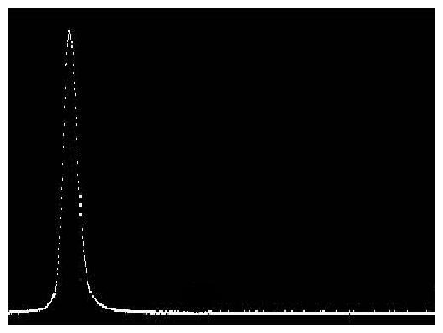}
      \begin{fig}\label{multicanale}
      \begin{center}
Immagine del picco di coincidenze restituito dal multicanale.
      \end{center}
      \end{fig}
\end{center}
\end{figure}

Per visualizzare e studiare 
questo aspetto dell'acquisizione dati,
si utilizza 
un multicanale
(Multi Channel Analizer,
indicato nel seguito con la sigla MCA)
costituito da 8196~canali,
riportati in ascissa,
ognuno dei quali è proporzionale
alla tensione del segnale inviatogli dal~TAC
e quindi al ritardo che caratterizza
la cattura delle coppie di fotoni.\\
L'intervallo temporale coperto dal multicanale è di 20~$ns$,
valore impostato 
mediante il~TAC.\\
Nel caso ideale dell'esempio precedente,
ciò che si osserva è una delta di Dirac,
poichè c'è sempre un solo canale che si attiva,
quello proporzionale al ritardo che caratterizza l'acquisizione.\\
Se invece si ha una dispersione dei ritardi si attivano anche
gli altri
e l'immagine restituita dall'MCA~(fig.~\ref{multicanale})
è un istogramma in cui in ordinata
è riportato il numero di coppie 
rilevate per canale.\\
L'FWHM indica di quanto ci si discosta dal
caso ottimale della 
figura piccata a forma di
delta di Dirac.\\

Per selezionare il picco,
relativo agli impulsi corrispondenti
ad eventi correlati, che sull'istogramma 
può apparire più o meno evidente occorre utilizzare un
Single Channel Analizer, SCA,
il quale, selezionata una sotto finestra interna a quella dei 20~$ns$
coperta dal~TAC e dal multicanale,
conta il numero di coppie 
catturate con un dato ritardo e 
corrispondenti al 
gruppo di canali prescelto.\\
Le osservabili che si intendono misurare
nell'esperimento sono
le concidenze,
ovvero il numero di fotoni rilevati simultaneamente
(entro una finestra temporale impostata sul TAC-SCA)
dai due fotorivelatori, e i conteggi relativi
ad ogni singolo dispositivo.
Infatti, la produzione di coppie di fotoni entangled
per~PDC avviene su scale temporali
dell'ordine del femtosecondo,
quindi la loro detezione genera delle coincidenze,
alle quali  si aggiungono quelle 
dovute a fotoni non correlati
che possono però essere valutate e sottratte a
quelle totali.\\
L'importanza di individuare solo le coincidenze dovute
a fotoni correlati è fondamentale
per i fini del nostro esperimento
come verrà illustrato nel capitolo~\ref{dBB}.\\
I segnali in uscita dallo~SCA  e
dai due rilevatori
sono successivamente inviati 
a dei contatori di un quad-counter
per monitorare la scansione.\\
Il quad-counter è un dispositivo 
costituito da quattro contatori
(il primo è un timer,
due servono per i singoli
rilevatori ed uno per le coincidenze) 
che può essere gestito anche in remoto
(in alcune fasi dell'esperimento si è
seguita questa via).\\
Ci sono due modi per realizzare ciò:
si può usare l'interfaccia~IEEE-488
oppure un cavo di tipo~RS-232-C
tramite porta seriale.
Per esigenze sperimentali si è 
adottata questa seconda metodologia
che ha richiesto la scrittura di un 
programma in linguaggio
pascal (riportato in appendice~\ref{quad.pas}).\\
Tale software permette di avere un'interfaccia 
grafica al calcolatore mediante la quale
è possibile digitare i comandi in linguaggio mnemonico
per operare a distanza sul quad-counter.\\
Per permettere una corretta comunicazione
tra calcolatore e contatore è stato necessario 
configurare gli switch di questo dispositivo
in maniera opportuna.\\
Il linguaggio pascal è stato usato
anche per scrivere un programma 
che gestisce la movimentazione
elettronica dei rilevatori
durante una scansione.\\
Infatti, è fondamentale il corretto posizionamento
dei dispositivi di detezione,
ciò viene realizzato mediante sistemi
di micromovimentazione manuali e
motori pilotati dal computer.
Nel nostro caso sono presenti
due carrelli per la traslazione dei
rilevatori e un solo rotatore,
tutto ciò è sufficiente a gestire una scansione.
Questo tipo di presa dati consiste nel
traslare e ruotare di un angolo opportuno
un rilevatore, per raggiungere la
posizione voluta, mentre l'altro
rimane fermo;
in questa configurazione si acquisiscono i dati,
quindi, si passa successivamente
ad un'altra posizione.\\
Il programma 
(riportato in appendice~\ref{looptest.pas})
ripete questa procedura il numero di volte richiesto
e, alla fine di tale operazione,
riporta i due elementi nelle posizioni iniziali.

\begin{figure}[h]
\includegraphics[width=14cm]{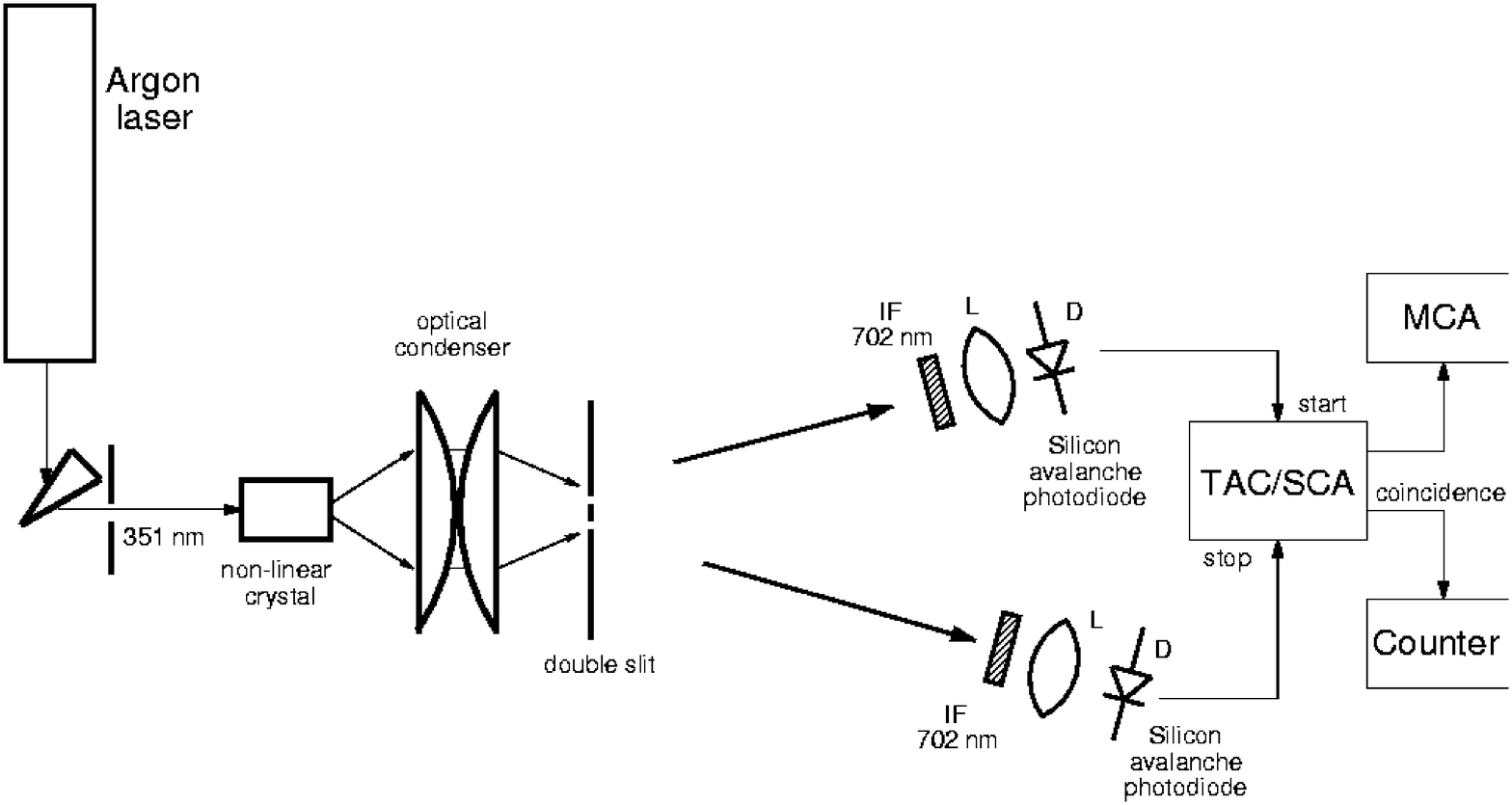}
      \begin{fig}\label{schema}
Schema della configurazione sperimentale adottata per la
nostra esperienza.
      \end{fig}
\end{figure}

In figura~\ref{schema} si riporta 
uno schema sintetico che descrive la configurazione
strumentale adottata per questo esperimento.

\clearpage

\section {Allinemento con fasci a 633~$nm$ e 789~$nm$}

Descriviamo, ora, la procedura iniziale di allineamento
realizzata mediante l'uso di emissioni parametriche 
a 633~$nm$ e 789~$nm$.
L'esperimento richiede l'uso di
fotoni con un lunghezza d'onda 702~$nm$, 
ma i fasci con tale lunghezza d'onda,
emessi dal cristallo di~$LiO_3$, 
sono di difficile individuazione pratica.\\
Al fine di avere dei punti di riferimento visibili 
si è iniettato nel cristallo di~$LiO_3$,
oltre alla pompa,
un secondo fascio 
di un laser a diodo a~789~$nm$
nella stessa direzione dell'emissione 
spontanea a 789~$nm$,
il quale induce un'emissione stimolata
a~633~$nm$.

\begin{figure}[h]
\begin{center}
\includegraphics[width=4cm]{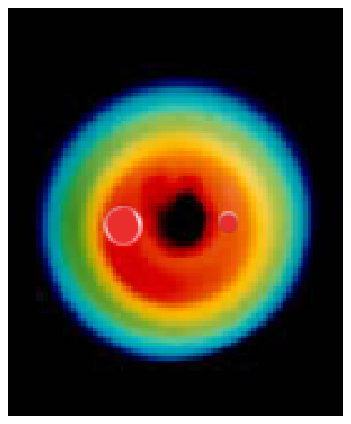}
      \begin{fig}\label{pdcstimolata}
      \begin{center}
Fotografia della PDC di tipi~I con emissione stimolata.
Lo spot più intenso (a sinistra) è dovuto al fascio del
laser a diodo a~789~$nm$ iniettato nel cristallo,
il meno intenso (a destra) è l'emissione stimolata
a~633~$nm$.
      \end{center}
      \end{fig}
      \end{center}
\end{figure}

Le due emissioni (633~$nm$ e 789~$nm$) 
permettono di identificare
un piano sul quale giacciono, non solo, le traiettorie 
dei fotoni con lunghezza d'onda 789~$nm$ e 633~$nm$,
ma anche quelle con $\lambda$=702$nm$, 
dunque, di fissare in prima
approssimazione la quota dei fotorivelatori.
Tale tecnica consente, inoltre,
di studiare l'efficienza
quantica strumentale con la quale tarare
l'intero apparato di rilevazione.

\section{Taratura dell'apparato sperimentale}\label{taratura}

L'efficienza dichiarata dei rilevatori è del 70~\%
nella regione attorno a  700~$nm$,
ma le ottiche utilizzate diminuiscono tale valore,
in particolare è interessante sapere come
modifichino l'acquisizione dati la presenza delle 
lenti con rivestimento anti-riflesso a 702~$nm$ 
(introdotte al fine di allargare
l'angolo solido intercettato dai fotorivelatore)
ed i filtri (necessari per diminuire il segnale
di fondo) per selezionare i fotoni a questa lunghezza d'onda.\\
Per questo motivo si è proceduto con una 
taratura dell'intero apparato 
utilizzando una tecnica consolidata in
applicazioni metrologiche~\cite{taratura}.\\

\subsection{Descrizione.}

\begin{figure}[h]
\begin{center}
\includegraphics[width=10cm]{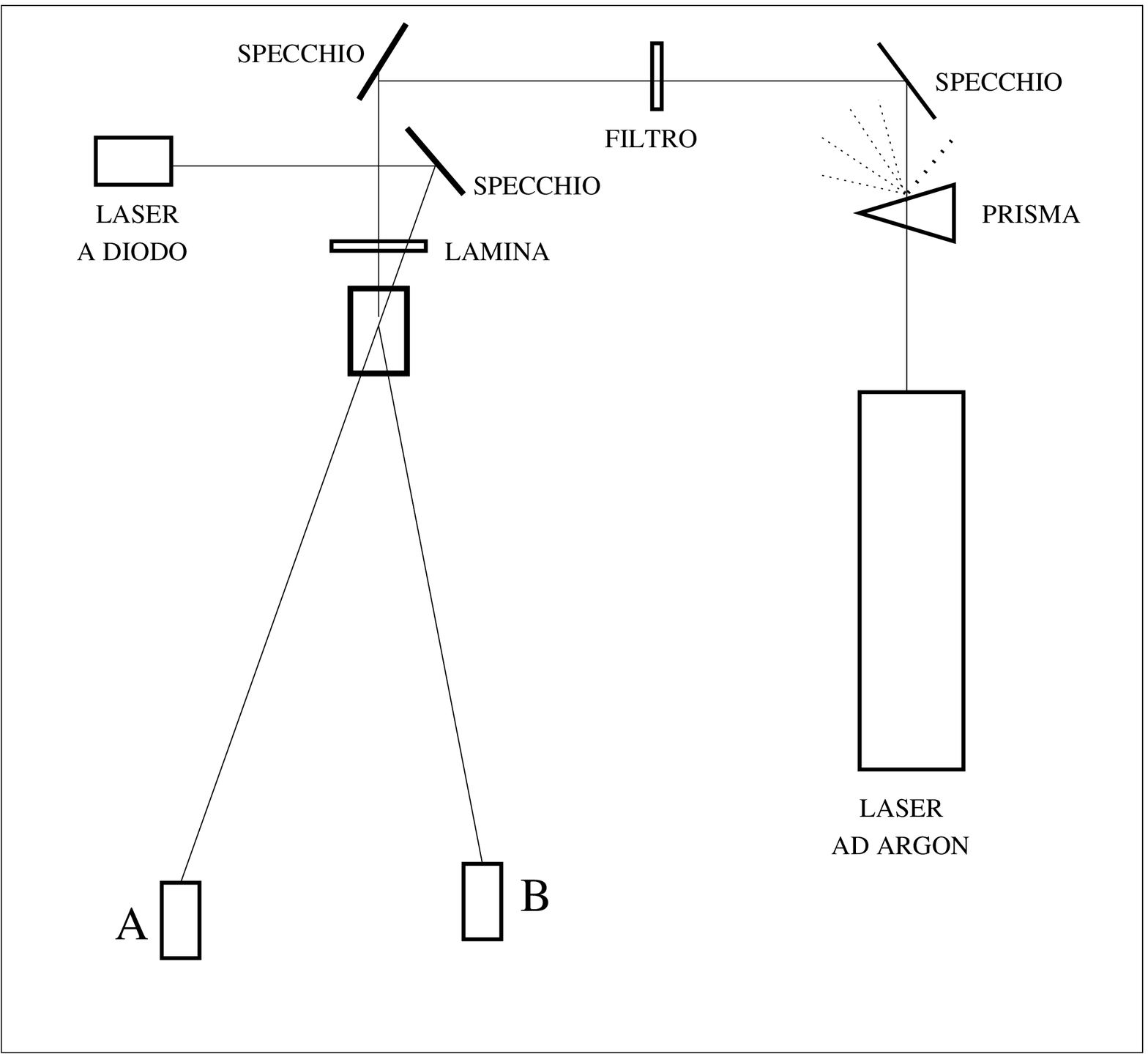}
      \begin{fig}\label{schemataratura}
Schema della configurazione sperimentale adottata per la taratura
dell'apparato di rilevazione.
      \end{fig}
\end{center}
\end{figure}

Al fine di allineare le ottiche si effettua una
selezione di due direzioni correlate iniettando
nel cristallo, con  un angolo opportuno, 
oltre al fascio di pompa del laser ad Argon,
un secondo fascio generato da
un laser a diodo a 789~$nm$
il quale origina un'emissione stimolata
a 633~$nm$ (fig.~\ref{schemataratura}).\\
I rilevatori che sono posti di fronte al cristallo
(vedi figura~\ref{schemataratura})
intercettano, rispettivamente, il fascio con lunghezza d'onda
789~$nm$ (il sinistro, denominato~A) e
633~$nm$ (il destro, indicato con la lettera~B).\\
Su dispositivi solidali con i fotorivelatori sono stati montati
filtri interferenziali centrati su tali  lunghezze d'onda: 
questi attenuano del~40\% la radiazione incidente in corrispondenza
della lunghezza d'onda prescelta 
(portando quindi ad una diminuzione dell'efficienza quantica
strumentale), ma di 3~ordini di grandezza quelli dovuti
ad altre lunghezze d'onda che creerebbero rumore di fondo
nell'acquisizione dati.\\

\subsection{Taratura.}\label{paragrafotaratura}

L'efficienza $\eta$ dell'intero sistema di rilevazione, 
in prima analisi, è definita come il rapporto
tra il numero di eventi registrati e il numero di
fotoni incidenti.
Grazie alle proprietà di correlazione della PDC
il flusso $\Phi$ di fotoni sui due canali
è identico.
Il numero di eventi registrati in un intervallo di tempo $\tau$
risulta, quindi, essere:

\begin{equation}\label{N_A}
N_A= \eta_A \Phi
\end{equation}

\begin{equation}\label{N_B}
N_B= \eta_B \Phi
\end{equation}

\begin{equation}\label{N_C}
N_C= \eta_A \eta_B \Phi
\end{equation}

Da cui segue:

\begin{equation}\label{efficienzaA}
\eta_A=\frac{N_C}{N_B}
\end{equation}

\begin{equation}\label{efficienzaB}
\eta_B=\frac{N_C}{N_A}
\end{equation}

dove $N_A$, $N_B$ e $N_C$
sono, rispettivamente, i conteggi
dei fotorivelatori~A e~B e delle coincidenze.\\
La conoscenza di $N_A$, $N_B$ e $N_C$
permette, dunque, 
di ottenere l'efficienza quantica dei due rilevatori,
senza dover valutare i flussi dei fotoni.
Si tratta, quindi, di un metodo di taratura
assoluto, indipendente dalla conoscenza
del flusso emesso da una sorgente di riferimento.\\
Un ulteriore vantaggio di tale metodologia è dovuto
al fatto che le misure non sono limitate
a particolari lunghezze d'onda,
dal momento che ogni possibile coppia di
fotoni correlati, nell'intero spettro della PDC, 
presenta le medesime caratteristiche di correlazione.

Per misurare l'efficienza quantica di un rilevatore 
(e dell'intero sistema) occorre usare il secondo come trigger,
ad esempio, per valutare $\eta_A$ si utilizza il fotorivelatore~$B$
come trigger.
Su tale apparato bisogna effettuare la massima selezione,
mediante filtri e diaframmi, in moto tale da considerare i 
conteggi provenienti da questo come dovuti ai soli fotoni correlati
riducendo il più possibile il contributo di fondo.\\
Le coincidenze che la catena elettronica è in grado di
mettere in rilievo con questa configurazione
sono quindi dovute a coppie di fotoni correlati temporalmente
visti da entrambi i fotorivelatori.
Tale numero, in una situazione ideale, dovrebbe coincidere
con i conteggi del trigger, ma ciò in pratica non si verifica
perché alcune particelle della coppia rilevate
da quest'ultimo sono state perse dallo
strumento che è oggetto di taratura.
In seguito a ciò  l'efficienza quantica,
che in prima approssimazione è stimata dal rapporto
tra misure delle coincidenze e dei conteggi del trigger ($N_A$),
sarà, quindi,
minore di~1:

\begin{equation}\label{efficienzaminorediuno}
\eta_B=\frac{N_C}{N_A}<1
\end{equation}

A questa prima stima dell'efficienza occorre
apportare alcune correzioni per tenere conto 
del rumore di fondo prodotto da
coincidenze casuali.\\
Dato il numero relativamente elevato di fotoni incidenti
la probabilità statistica che due di essi, 
pur non essendo prodotti in coppia all'interno del cristallo, 
vengano rilevati simultaneamente 
(cioè dentro una finestra temporale di 20 ns) è non nulla.\\
Ci si attende infatti un contributo dato da:
\begin{equation}
N_{c.f.p.}=N_{A} \cdot N_{B} \cdot t
\label{Coincidenze statistiche}
\end{equation}

Dove~$N_{c.f.p.}$ sono i  {\bfseries c}onteggi {\bfseries f}uori dal {\bfseries p}icco,
$N_{A}$~sono quelli relativi ad~A e
$N_{B}$~quelli di~B;
t~è la finestra temporale di acquisizione determinata dal 
TAC~(5~ns).\\
Al fine di stimare le coincidenze accidentali nel picco,
si assume che tali valori siano costanti, quindi,
si procede con un'acquisizione in un intervallo
temporale traslato rispetto al picco e di pari durata.
Tale traslazione temporale si ottiene modificando 
il ritardo nella catena elettronica di
modo che lo SCA non sia più centrato sul picco,
ma vada a contare le coincidenze in un'altra regione.\\
Inoltre, all'interno del cristallo 
si possono verificare fenomeni di fluorescenza 
o diffusione del laser di pompa
che contribuiscono al segnale di fondo del rilevatore
di trigger.
Per individuare questo contributo si introduce una lamina a quarto d'onda 
(nella posizione indicata in figura~\ref{schemataratura})
la quale ruota di~$\frac{\pi}{2}$ la polarizzazione del laser: 
sopprime così la~PDC, 
ma resta il contributo di fluorescenza e diffusione del laser di pompa.
I conteggi che si rilevano sono, quindi, 
dovuti a tali componenti.\\
In realtà la situazione non è così ottimale: 
resta sempre una piccola componente trasversa che 
origina fluorescenza parametrica 
e quindi si continua a vedere un picco di coincidenze relativo alla 
produzione di coppie all'interno del cristallo, 
come può essere mostrato dall'immagine restituita dal
multicanale.
Tuttavia, questa componente può essere trascurata in una 
prima approssimazione.\\
Introdotte dette correzioni,
l'efficienza quantica è quindi data da:

\begin{equation}
\eta = \frac{N_{c}-N_{c.f.p.}}{N_{t}-N_{l}}
\label{efficienza}
\end{equation}

Dove~$N_c$ sono i conteggi relativi alle coincidenze,
~$N_{c.f.p.}$ le coincidenze fuori dal picco,
~$N_t$ i conteggi di singolo canale relativi al fotorivelatore
utilizzato come trigger e  $N_l$ quelli presi dopo 
aver posizionato sul cammino del fascio di pompa
un lamina a quarto d'onda.\\
Per ottenere una stima corretta dell'efficienza quantica,
devono essere apportate ulteriori correzioni.\\
Innanzitutto si introduce in $\eta$ un fattore
$\frac{1}{T_{signal}}$, al fine di tenere conto
della trasmittanza ($T_{signal}$) del cammino
ottico dei fotoni del fascio {\parolestrane signal} 
che non è  ideale.\\
Può accadere, inoltre, che un evento dovuto ad una
coincidenza correlata non  venga registrato se l'impulso di stop
corrispondente all'evento correlato al segnale di start del trigger,
è preceduto ad un istante $t$
da un segnale di stop generato da un fotone scorrelato.\\
La frazione di conteggi mancanti è data da
($t_{delay}$ è il ritardo inserito sul segnale di stop):

\begin{equation}
\alpha \simeq 1- \int_{t_0}^{t_{delay}} W_{signal}(t')dt=%
1-\langle W_{signal} \rangle t_{delay}=%
1-\frac{N_{signal}}{T}t_{delay}
\end{equation}

dove $W_{signal}$ denota i conteggi sul canale 
{\parolestrane signal} e
$T$ è il tempo di una acquisizione.\\
Si introduce un fattore $\frac{1}{\alpha}$ che 
esprime la correlazione tra le coincidenze totali
ed accidentali.\\
Al fine di tenere conto del tempo morto del rilevatore oggetto di taratura
($\tau_{signal}$),
è necessario aggiungere un ulteriore fattore correttivo:

\begin{equation}
\gamma \simeq 1-N_{signal} \tau_{signal} \frac{1}{T}
\end{equation}

L'efficienza dell'apparato di rilevazione risulta, dunque, 
essere data dalla seguente espressione:

\begin{equation}
\eta_B=%
\frac{N_{c}-N_{c.f.p.}}{N_{t}-N_{l}}%
\frac{1}{T_{signal}}%
\frac{1}{\alpha}%
\frac{1}{\gamma}
\end{equation}

\subsection{Scansione.}

Al fine di tarare un rilevatore non è sufficiente
calcolare l'efficienza quantica
in un punto, ma occorre eseguire una scansione
cioè, mentre il trigger è fisso l'altro dispositivo viene
traslato con passi da 850 $\mu m$;
tale operazione è stata ripetuta per entrambi
i dispositivi di rilevazione.
Si identifica, così, il punto corrispondente 
alla massima efficienza,
ovvero al miglior allineamento.
Con tale operazione si valuta la superficie 
dell'apparto di  rilevazione
inoltre, si determina la quota del piano su cui giacciono
i fasci a 702~$nm$.

\subsection{Risultati.}

\begin{figure}[h]
\begin{center}
\includegraphics[width=15cm]{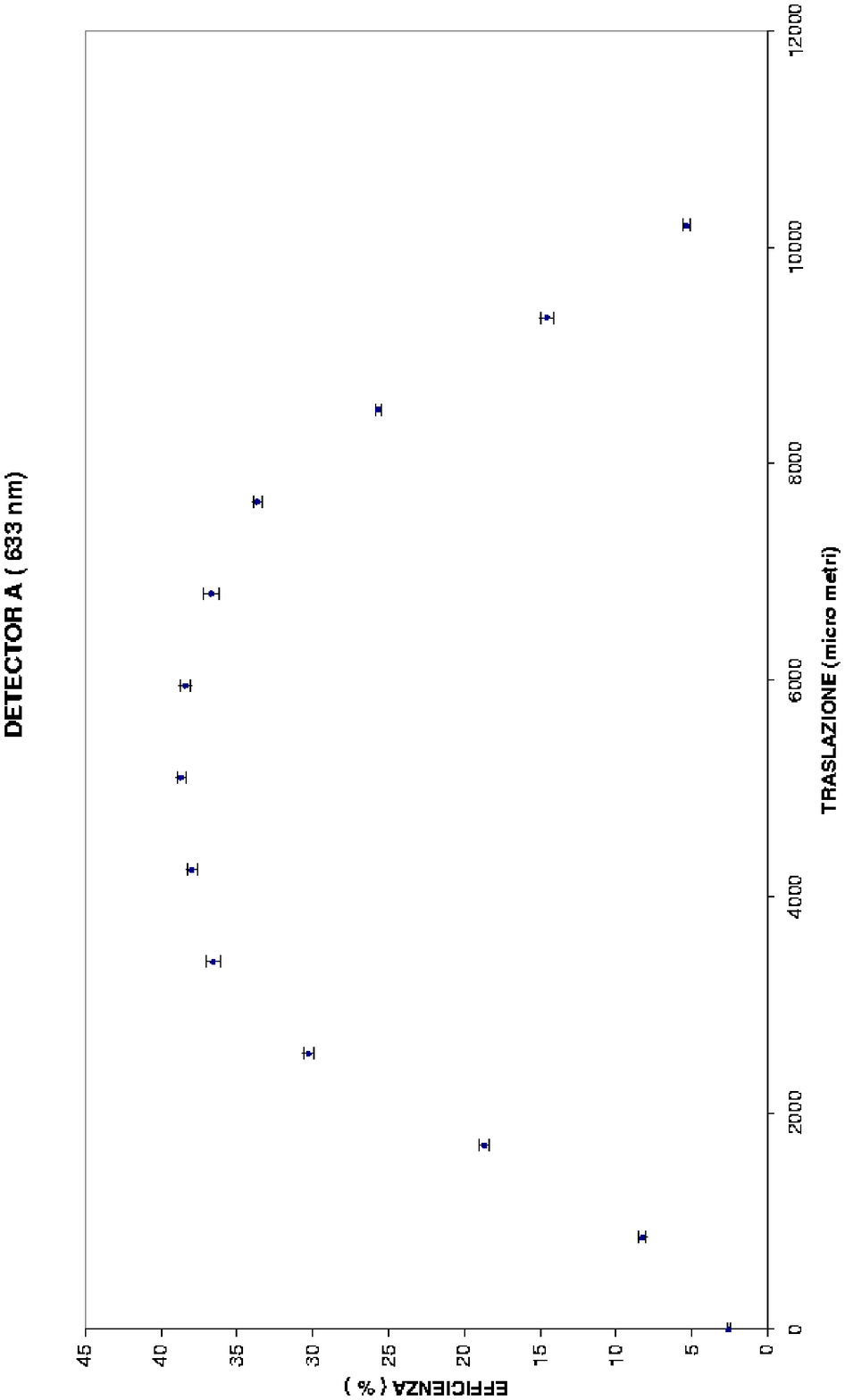}
      \begin{fig}\label{taratura633}
Curva di taratura del fotorivelatore con
lente con rivestimento anti-riflesso per 633~$mm$.
      \end{fig}
\end{center}
\end{figure}

\begin{figure}[h]
\begin{center}
\includegraphics[width=15cm]{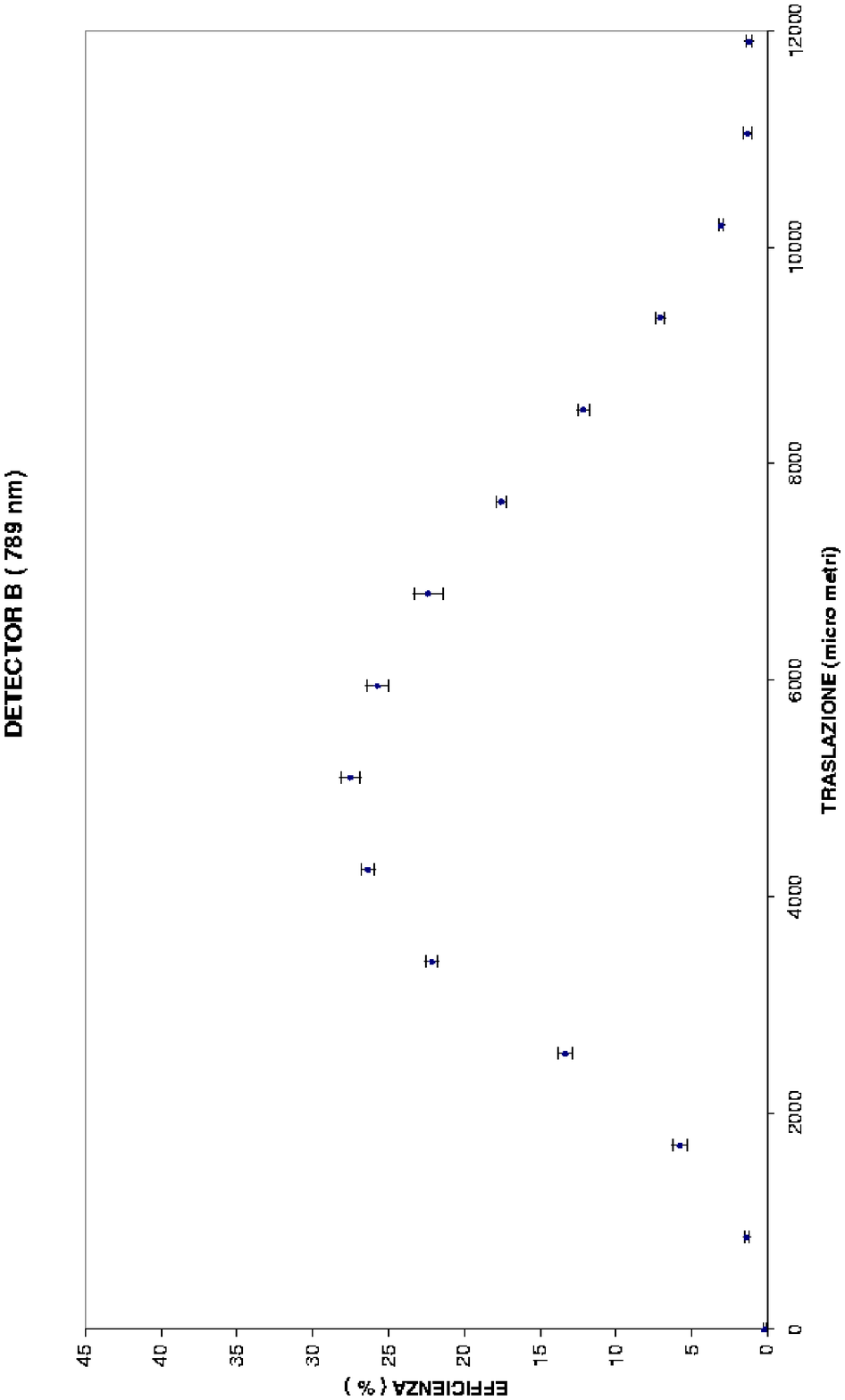}
      \begin{fig}\label{taratura789}
Curva di taratura del fotorivelatore con
lente con rivestimento anti-riflesso per 789~$mm$.
      \end{fig}
\end{center}
\end{figure}

I risultati ottenuti sono riportati in figura~\ref{taratura633} 
e~\ref{taratura789}.
E' possibile notare come 
l'introduzione delle lenti con rivestimento anti-riflesso abbia 
allargato la superficie di raccolta dell'apparato di rilevazione.\\
E' bene precisare che tali ottiche sono 
state studiate per fotoni con 
lunghezza d'onda 702~$nm$,
cioè quelli utilizzati per l'esperimento 
oggetto di studio della presente tesi,
e non per quelli a 633~$nm$ e 789~$nm$ 
usati in questo caso. 
Tuttavia dalle specifiche tecniche
emerge che la differenza è trascurabile,
inferiore al~1\%.\\
L'efficienza varia da punto a punto,
poichè il fascio inizialmente 
colpisce la lente su un bordo,
in cui si ha un valore di efficienza praticamente nullo,
per poi spostarsi verso il centro
dove si ha un picco massimo  del~$(27.5 \pm 0.7) \%$
per il~B
e~$(40.0 \pm 0.3) \%$
per~A.
Si può, quindi, concludere che
la superficie di detezione coincide con quella 
delle lenti, cioè ha un diametro di circa 6~$mm$.\\

\clearpage

\clearpage

\clearpage

\section{Allineamento con fasci a 702~$nm$}

\begin{figure}[h]
\begin{center}
\includegraphics[width=11cm]{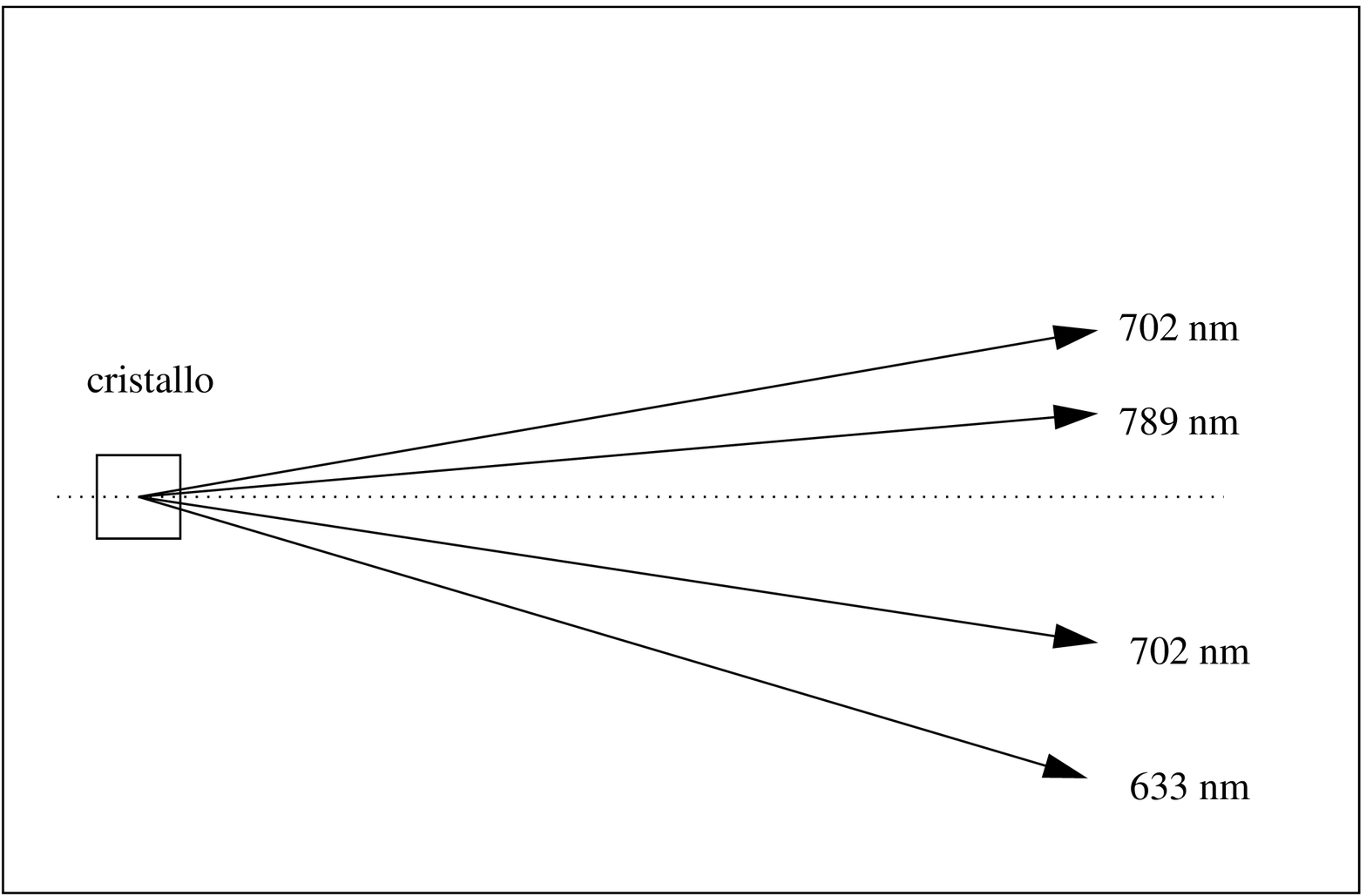}
      \begin{fig}\label{traiettorie}
Traiettorie delle coppie di fotoni entangled
emessi per PDC dal cristallo di $LiO_3$,
in cui si osserva come variano le direzioni in
funzione della lunghezza d'onda del quanto considerato.
      \end{fig}
\end{center}
\end{figure}

In questa seconda fase si procede,
quindi, con un allineamento a 702~$nm$,
dopo aver sostituito i 
filtri interferenziali a 633~$nm$ e 789~$nm$
posizionati di fronte agli apparati di rilevazione
con altri centrati sulla lunghezza d'onda 702~$nm$.\\
L'identificazione precisa delle direzioni 
correlate a 789~$nm$ e 633~$nm$ ha permesso
di posizionare i fotorivelatori alla quota 
del piano su cui giacciono le traiettorie
dei fotoni con lunghezza d'onda 702~$nm$;
noto l'angolo con cui tali fotoni sono emessi
dal cristallo di~$LiO_3$ è possibile orientarli
nella direzione voluta~(fig.~\ref{traiettorie}).\\
Si procede, successivamente, 
all'inserimento della doppia fenditura.

\clearpage

\section{Posizionamento della doppia fenditura}

Occorre, ora, inserire la doppia fenditura
in modo tale che ogni fotone a 702~$nm$ 
della coppia passi per una specifica apertura.
Al fine di verificarne il corretto posizionamento
si rimuove il cristallo di $LiO_3$
e si invia direttamente il fascio di pompa
verso le fenditure.
Chiudendo alternativamente una delle due aperture,
mediante l'uso di due lamette montate
su un sistema di micromovimentazione,
si osserva la figura di diffrazione fig.~\ref{frange_no}.\\
Ciò permette inoltre di determinare con precisione
la posizione delle lamine allorquando chiudono una
singola fenditura,
posizione che verrà utilizzata per
l'allineamento finale dell'esperimento.\\
Lasciando esposte entrambe le fenditure
alla radiazione di pompa (sufficientemente
larga da avere valore non nullo in prossimità
delle due fenditure)
osserviamo anche il contributo di interferenza
al secondo ordine~(fig.~\ref{frange_si}).

Effettuato questo primo allineamento della fenditura
si reinserisce il cristallo di
$LiO_3$ sul percorso del fascio di pompa,
generando fotoni entangled a 702~$nm$
che vengono indirizzati ognuno verso una
specifica fenditura.\\
Eliminato il laser di pompa con un filtro
le due fenditure vengono quindi
raggiunte dall'emissione correlata
a~702~$nm$.
Si osserva un picco di coincidenze
che viene massimizzato.
Si controlla quindi che le coppie 
di fotoni attraversino fenditure distinte
chiudendo alternativamente
una delle due fenditure
ed osservando la sparizione del picco.

\clearpage

\begin{figure}[h]
\includegraphics[width=14cm]{figure/frange_no.eps}
      \begin{fig}\label{frange_no}
Fotografia della figura di diffrazione 
(senza interferenza)
prodotta dal fascio laser di pompa
che incide solo su una delle due fenditure.
      \end{fig}
\end{figure}

\begin{figure}[h]
\includegraphics[width=14cm]{figure/frange_si.eps}
      \begin{fig}\label{frange_si}
Fotografia della figura di diffrazione ed interferenza 
prodotta dal fascio laser di pompa
che incide su entrambe le fenditure.
      \end{fig}
\end{figure}

\section{Previsioni quantistiche}

La funzione d'onda che descrive l'emissione parametrica
può, in forma semplificata, essere scritta nel seguente 
modo:

\begin{equation}
|\Psi\rangle =|vac\rangle +\int d\omega_i d\omega_s \Phi (\omega_i, \omega_s)|\omega_i\rangle |\omega_s\rangle
\end{equation}

\begin{figure}[h]
\includegraphics[width=14cm]{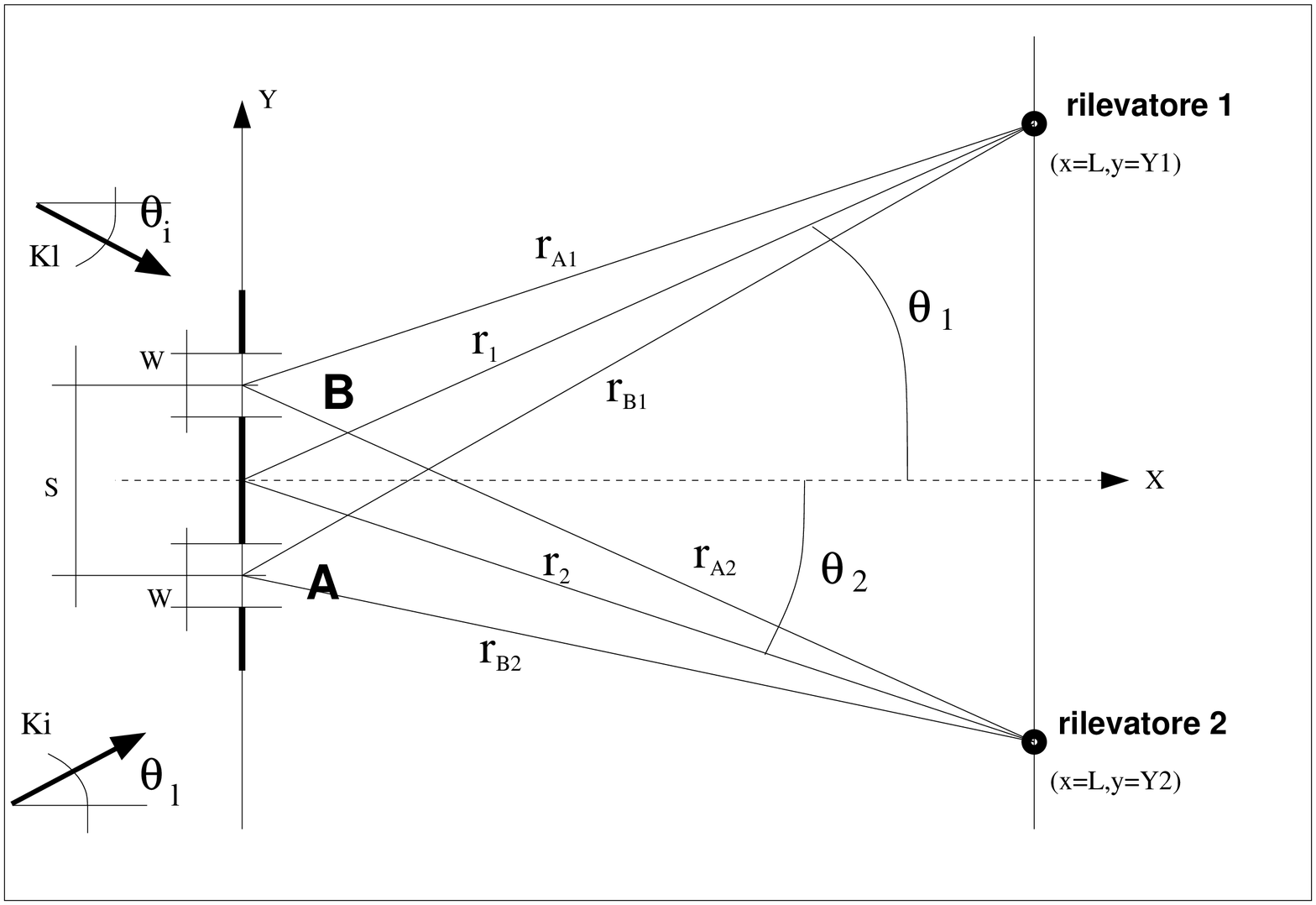}
      \begin{fig}\label{schemafunzione}
Schema geometrico relativo al modello quantistico standard
adottato per la propagazione della radiazione trasmessa
attarverso una doppia fenditura.
      \end{fig}
\end{figure}

Nella regione di Fraunhofer, cioè a grande distanza
dalla doppia fenditura,
(con riferimento alla figura~\ref{schemafunzione}),
il campo diffratto è descritto da:

\begin{equation}\label{funzionenellimitedif}
\Phi(\omega_i,\omega_s)=%
g(\theta_1,\theta_i^A)g(\theta_2,\theta_i^B)%
e^{-i(k r_{A1}+k r_{B2})}+%
g(\theta_2,\theta_i^A)g(\theta_1,\theta_i^B)%
e^{-i(k r_{A2}+k r_{B1})}
\end{equation}

\begin{equation}
g(\theta,\theta_i^l)=%
\frac{%
sin(kw/2(sin\theta -sin\theta_i^l))
}{%
kw/2(sin\theta - sin\theta_i^l)
}
\end{equation}

Dove $k$ è il vettore d'onda del fotone
A~({\parolestrane idler}) e del fotone B~({\parolestrane signal}),
$r_{ai}$ è il vettore che congiunge la fenditura 
$a$~(che può essere A o~B) col fotorivelatore~$i$ (1 o~2),
con $\theta_i^l$ si denota l'angolo di diffrazione del fotone
sulla fenditura~$l$ (A o~B),
$s$ è la separazione tra le due fenditure
di larghezza~$w$.
$g(\theta,\theta_i^l)$ sono definiti 
secondo l'ottica ondulatoria nel limite di Fraunhofer,
nell'ipotesi di una distribuzione di probabilità uniforme di trovare
un  fotone in un punto specifico della fenditura.
La funzione~$\Phi$ è simmetrica per scambio dei fotoni
essendo, essi, bosoni indistinguibili.
Dall'equazione~\ref{funzionenellimitedif}
si ottiene una relazione che esprime il numero di 
coincidenze atteso 
in funzione delle variabili $\theta_i^l$:

\begin{eqnarray}
\label{moduloquadrodellafunzionepsi}
C(\theta_1,\theta_2)=|\Phi(\omega_i,\omega_s)|^2=%
g(\theta_1,\theta_i^A)^2 g(\theta_2,\theta_i^B)^2+\nonumber\\%
g(\theta_2,\theta_i^A)^2 g(\theta_1,\theta_i^B)^2+%
2%
g(\theta_1,\theta_i^A)^2 g(\theta_2,\theta_i^B)^2*\nonumber\\%
g(\theta_2,\theta_i^B)^2 g(\theta_1,\theta_i^B)^2%
cos[ks(sin\theta_1-sin\theta_2)]
\end{eqnarray}

Essa mostra una figura d'interferenza con frange 
distanziate $\Delta y=L \frac{\lambda}{s}$,
dove L è la distanza tra la doppia fenditura
e l'asse $y$ lungo il quale sono disposti
i rilevatori.
Tali frange sono comprese tra picchi di intensità:

\begin{equation}
I_1=[%
g(\theta_1,\theta_i^A) g(\theta_2,\theta_i^B)-%
g(\theta_2,\theta_i^A) g(\theta_1,\theta_i^B)]^2
\end{equation}

\begin{equation}
I_2=[%
g(\theta_1,\theta_i^A) g(\theta_2,\theta_i^B)+%
g(\theta_2,\theta_i^A) g(\theta_1,\theta_i^B)]^2
\end{equation}

La diffrazione distribuisce la probabilità
di rilevazione congiunta lungo la direzione $y$.
L'andamento del $|\psi(\theta_1,\theta_2)|^2$
è riportato in fig.~\ref{graficomodulopsi2d}
in una rappresentazione bidimensionale e in 
fig.~\ref{graficomodulopsi3d} in una tridimensionale.\\
Al fine di valutare l'effetto della non monocromaticità
della radiazione si è calcolata la convoluzione 
della funzione~$|\psi|^2$ (nell'ipotesi 
che essa dipenda dalla lunghezza
d'onda di uno dei due fotoni correlati)
con la funzione di trasferimento gaussiana di un filtro interferenziale,
centrata su una lunghezza d'onda di 702~nm con FWHM pari a 20~nm.\\
Per filtri interferenziali con FWHM di 4~$nm$
tale correzione è risultata trascurabile.

\clearpage

\begin{figure}[h]
\includegraphics[width=16cm]{figure/graficomodulopsi2d.eps}
      \begin{fig}\label{graficomodulopsi2d}
Mappa bidimensionale della densità di probabilità 
di rilevazione congiunta di una coppia di fotoni
entangled in funzione delle posizioni dei rilevatori 
secondo il modello quantistico standard.
Le zone in rosso corrispondono ai massimi,
quelle in blu ai minimi.
      \end{fig}
\end{figure}

\clearpage

\begin{figure}[h]
\includegraphics[width=16cm]{figure/graficomodulopsi3d.eps}
      \begin{fig}\label{graficomodulopsi3d}
Rappresentazione tridimensionale della densità di probabilità
di rilevazione congiunta di una coppia di fotoni entangled
in funzione delle posizioni dei rilevatori 
secondo il modello quantistico standard.
      \end{fig}
\end{figure}

\clearpage

\section{Acquisizione ed analisi dati}\label{acquisizioneedanalisidati}

\begin{figure}[h]
\begin{centering}
\includegraphics[width=14cm]{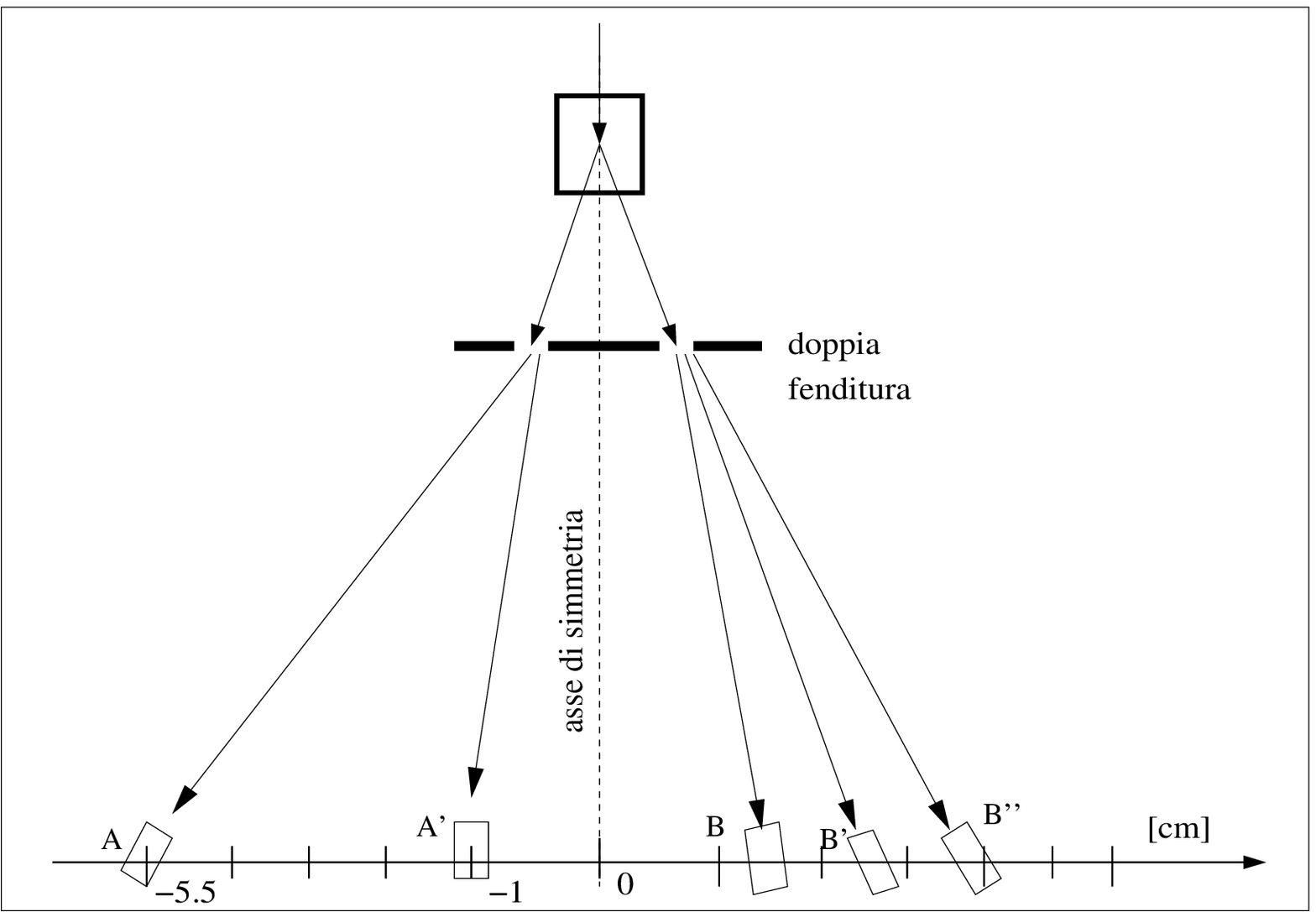}
      \begin{fig}\label{schemaasse}
Schema indicativo della posizione dei fotorivelatori~(A e B)
in funzione dell'asse di simmetria della doppia fenditura,
il quale
individua la posizione zero
degli apparati di rilevazione.
Le posizioni a sinistra di tale asse hanno segno
negativo, positive quelle a destra.
      \end{fig}
\end{centering}
\end{figure}

Con riferimento alla figura~\ref{schemaasse},
i fotorivelatori sono indicati con le lettere A e B,
gli apici denotano posizioni successive.\\
L'asse di simmetria della doppia fenditura
individua la posizione zero
degli apparati di rilevazione
e
divide in due parti il piano su cui giacciono
i fasci con lunghezza d'onda~702~$nm$
utilizzati per l'esperimento.
Le posizioni a sinistra di tale asse hanno segno
negativo, quelle a destra sono positive.\\
La scansione della figura di interferenza è 
realizzata mantenendo il dispositivo di rilevazione
di sinistra fisso 
e traslando il secondo 
di una distanza che dipende dal dettaglio 
con cui si vuole effettuare l'analisi,
la rotazione necessaria al fine di
mantenere il rilevatore puntato verso il
cristallo è calcolata in modo 
opportuno.

Al fine di tenere conto di un significativo 
rumore di fondo (dovuto a coincidenze casuali)
per le coincidenze si è considerato
il loro valore 
{\parolestrane `effettivo'}
valutato come differenza tra i dati 
acquisiti quando la finestra temporale
del TAC-SCA è centrata sul picco
e quelli presi ritardando la 
catena elettronica di misurazione di 16~ns.

\subsection{Fenomeno di deriva}

In fase di acquisizione dati si è osservato un fenomeno di
{\parolestrane deriva}:
pur non modificando nulla nella configurazione
sperimentale, cioè lasciando invariata la posizione
dei fotorivelatori e l'orientazione del cristallo,
si nota una variazione del numero di conteggi
su singolo canale e relativo alle coincidenze.
Le cause possono essere di diversa natura:
la potenza del laser dopo molte ore di lavoro
può diminuire; 
la polarizzazione del fascio prodotto può
cambiare  nel tempo;
il cristallo di $LiO_3$, 
in cui viene iniettato il laser può deteriorarsi
e quindi  possono variare  
le sue caratteristiche fisiche
da cui dipende la fluorescenza parametrica.\\
A causa della deriva del laser non si graficano le
coincidenze rilevate in funzione della posizione
dei rilevatore, ma quelle che definiamo
{\parolestrane coincidenze corrette} valutate 
moltiplicando il rapporto tra le
coincidenze effettive e i conteggi
effettivi di A per la media di tutti i
conteggi effettivi,
rilevati durante l'intera scansione,
dello stesso rilevatore.\\

\subsection{Analisi della figura di diffrazione e di interferenza.}

Lo studio della figura di interferenza e di
diffrazione prodotta dai fotoni 
si è articolato in tre fasi.

In una prima fase il rilevatore di sinistra 
è stato posizionato a -5.5~cm
(con riferimento alla 
figura~\ref{schemaasse}), 
mentre il destro è stato traslato con passi da
1~cm e ruotato di un angolo opportunamente valutato.
Raggiunta la posizione voluta si è proceduto con la presa
dati, della durata di mezz'ora, 
sia per il segnale che per il fondo.
Tali valori sono stati riportati sul grafico
in fig.~\ref{pattern1},
e sono riferiti ad un tempo di acquisizione 
di 10 minuti.
L'incertezza sullle
posizioni è stata valutata pari a 6~mm,
sulla base di quanto verificato durante 
la taratura dell'apparato sperimentale.\\
La curva che riproduce le previsioni teoriche 
della meccanica quantistica standard 
viene normalizzata ai punti sperimentali.
Il grafico riportato in fig.~\ref{pattern1}
mostra l'accordo  esistente tra le 
previsioni della MQS
e l'esperimento.
In questa configurazione sperimentale
non è possibile apprezzare il contributo
dovuto all'interferenza a causa dell'iride 
troppo larga rispetto al
periodo dell'interferenza stessa.
Il confronto di
questi dati con il calcolo svolto
da P.Ghose~\cite{propostaesperimentoghose}
verrà discusso nel capitolo~\ref{dBB}.

In una seconda fase si è posizionato il fotorivelatore di sinistra
a -1~cm~(fig.~\ref{schemaasse})
e si è ripetuta la stessa procedura già adottata nella
precedente fase, con passi da 1~cm ed acquisizioni
da 1~ora.
I dati sperimentali (riportati in fig.~\ref{pattern2})
mostrano un perfetto accordo con le previsioni quantistiche,
ma la configurazione sperimentale non consente
ancora una valutazione dettagliata 
delle frange di interferenza al quarto ordine.\\
Con riferimento alla simulazione riportata in 
fig.~\ref{piani} lo studio condotto nella prima fase
equivale ad una analisi del profilo della curva
in fig.~\ref{graficomodulopsi3d}
ottenuta sezionando la figura con un piano
passante per l'asse~I, nella seconda fase
il piano passa per l'asse~II.

Nell'ultima fase è stata aggiunta un'iride,
del diametro di 2~$mm$ di fronte ai rilevatori
al fine di ridurre l'angolo solido intercettato,
consentendo una selezione spaziale tale 
da permetterre un'analisi dettagliata 
della figura di interferenza al quarto ordine
(fig.~\ref{pattern3}).

\begin{figure}[h]
\begin{center}
\includegraphics[width=14cm]{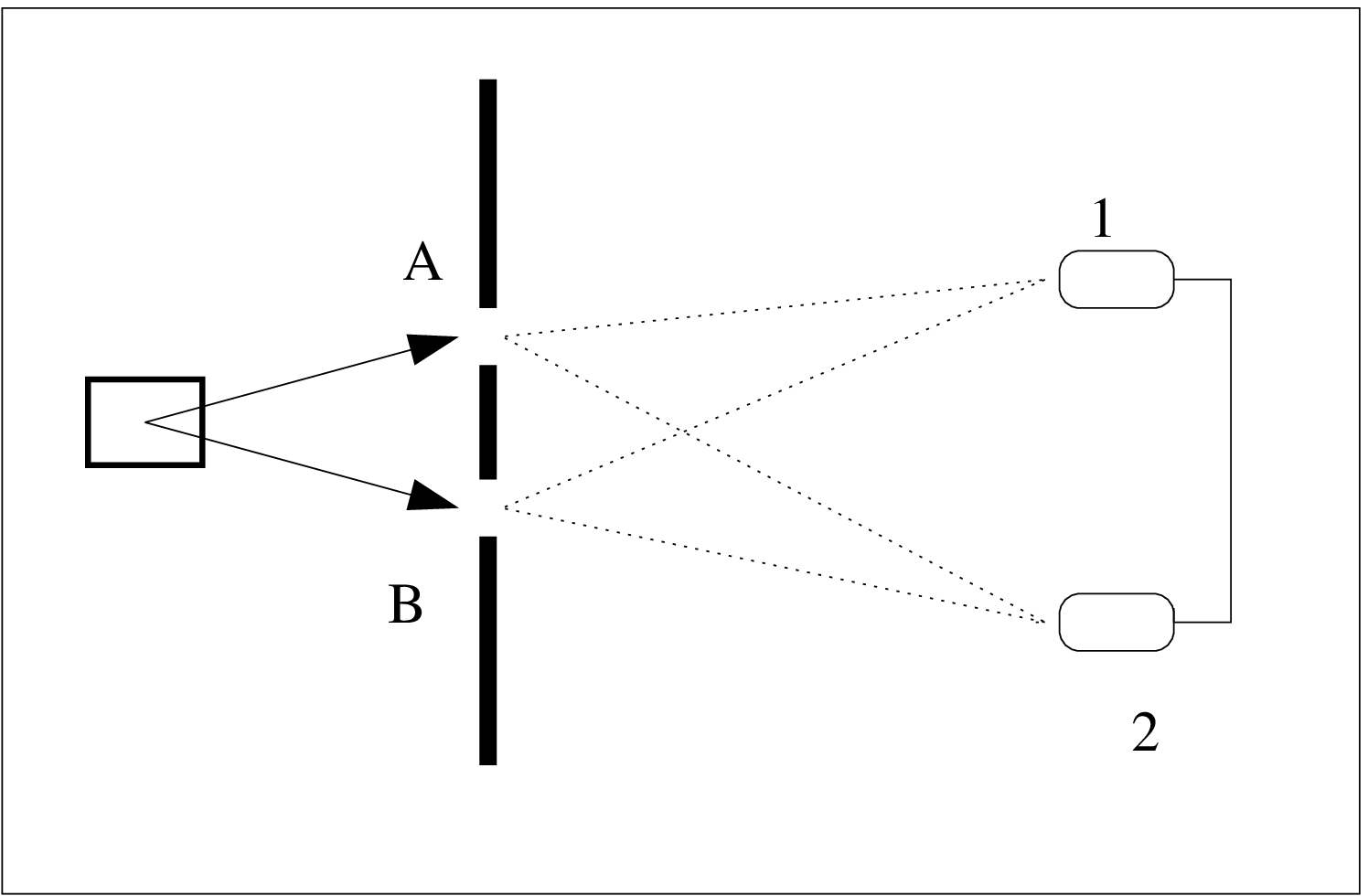}
\end{center}
       \begin{fig}\label{schemadetector2}
Schema della configurazione sperimentale, ove per convenzione 
si indicano con le lettere~A e~B le aperture della 
doppia fenditura e con~1 e~2 i fotorivelatori.
       \end{fig}
\end{figure}

\begin{figure}
\includegraphics[width=14cm]{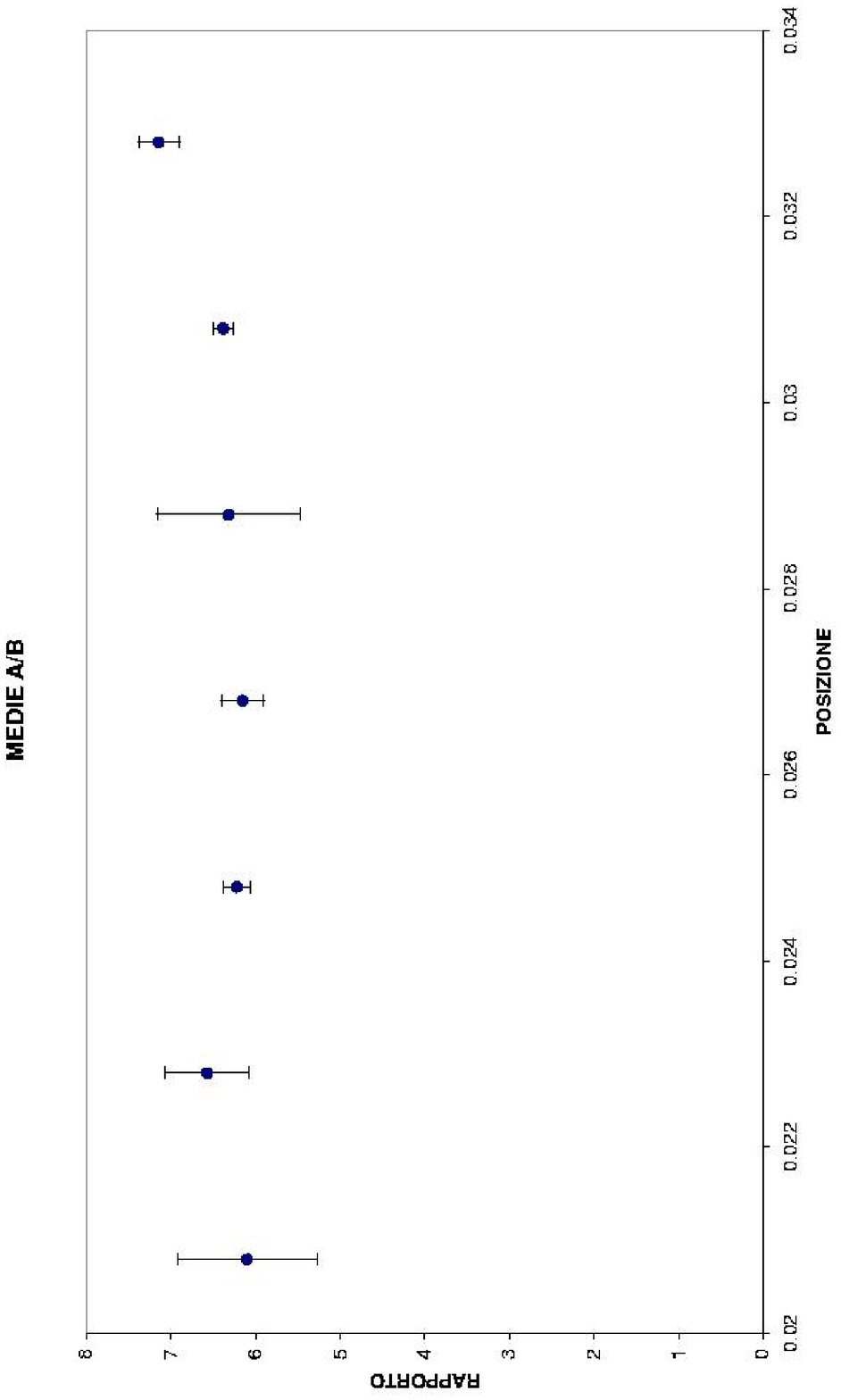}
      \begin{fig}\label{rapporto}
Rapporto dei conteggi
di singolo canale in funzione della distanza reciproca
dei due fotorivelatori.
L'andamento sostanzialmente costante di tali valori mostra 
un'assenza del fenomeno dell'interferenza al secondo
ordine.
      \end{fig}
\end{figure}

Come dimostrazione del fatto che si è realizzata
la configurazione desiderata,
in figura~\ref{rapporto} è riportato
il rapporto dei conteggi
di singolo canale in funzione della distanza reciproca
dei due fotorivelatori.
Tale rapporto è, come atteso, costante e non
mostra quindi interferenza al secondo ordine,
ovvero, a livello di singolo fotone 
poichè la traiettoria del singolo
quanto è perfettamente identificata
(la fenditura attraversata è nota).\\
D'altro canto l'interferenza al quarto ordine,
è chiaramente osservata (fig.~\ref{pattern3}).
Essa è dovuta al pacchetto bifotonico nel
suo complesso ed è
valutata mediante le coincidenze
prodotte dai fotoni correlati, ma 
a questo livello non è possibile determinare,
con riferimento alla figura~\ref{schemadetector2},
se il fotone rilevato da un dispositivo (1 o 2)
abbia attraversato una fenditura
(A o B) o l'altra.\\
Il risultato così ottenuto 
è una dimostrazione del fatto che 
i campi bifotonici debbano essere considerati 
nel loro complesso e non come due singoli fotoni.\\
La figura di interferenza prevista dalle meccanica
quantistica si adatta perfettamente ai dati sperimentali
restituendo un $\chi^2$ ridotto di~0.9.\\

\clearpage

\begin{figure}[h]
\includegraphics[width=14cm]{figure/piani.eps}
      \begin{fig}\label{piani}
Lo studio condotto nella prima fase della nostra esperienza
equivale ad una analisi del profilo della 
rappresentazione tridimensionale della densità di probabilità
di rilevazione congiunta di una coppia di fotoni entangled
(riportato in fig.~\ref{graficomodulopsi3d})
ottenuto sezionando la figura con un piano
passante per l'asse~I.
Nella seconda fase, tale piano passa per l'asse~II.
      \end{fig}
\end{figure}

\clearpage

\begin{figure}
\includegraphics[width=13cm]{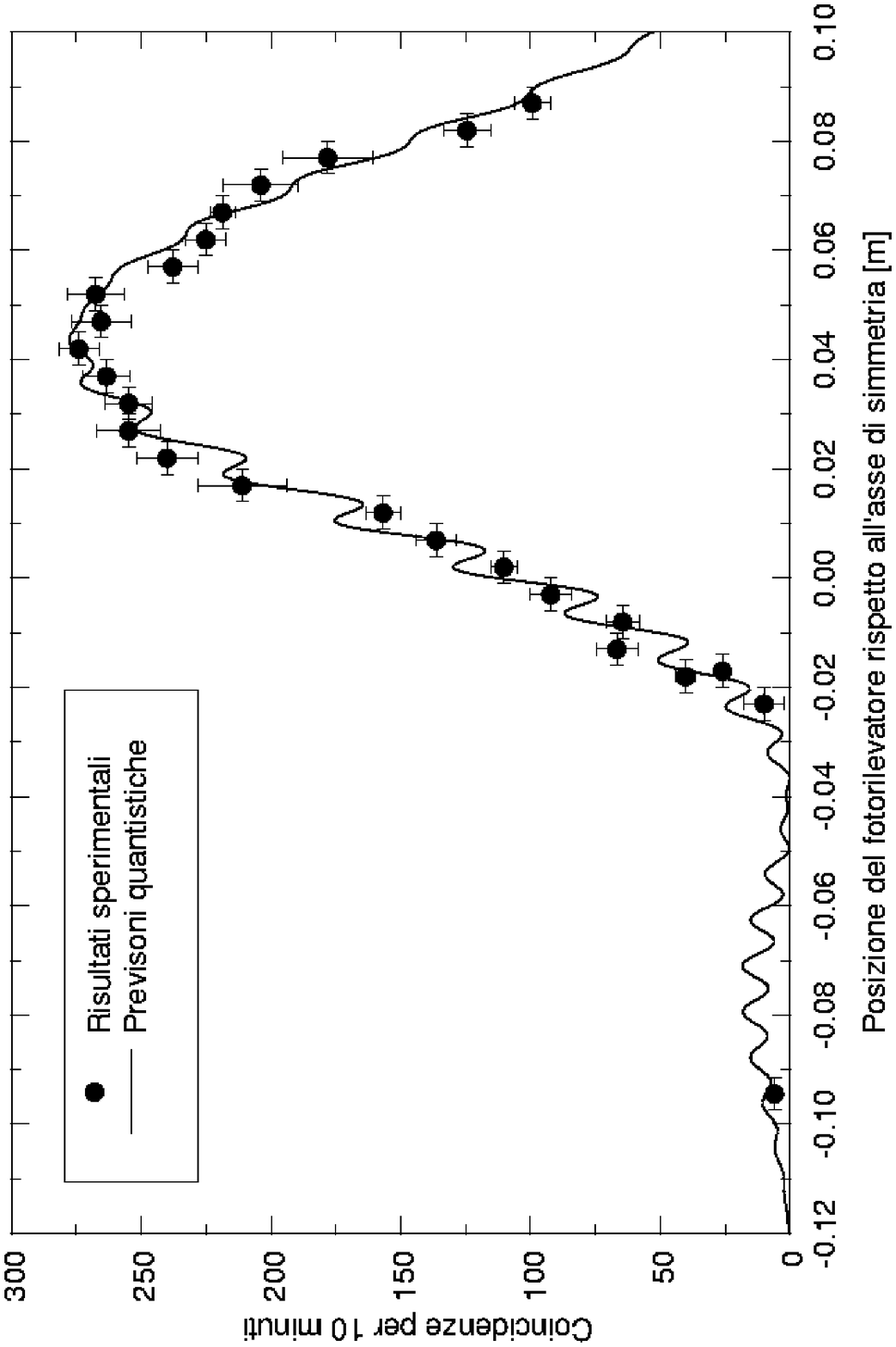}
      \begin{fig}\label{pattern1}
Profilo della figura di diffrazione prodotta 
dal passaggio di due fotoni entangled, 
di lunghezza d'onda 702~$nm$, 
attraverso una doppia fenditura,
ognuno per una determinata apertura.\\
Le previsioni quantistiche prevedono la
presenza di un picco principale e di 
uno secondario.
Tale profilo può essere ricavato
dalla figura~\ref{graficomodulopsi3d}
sezionando con il piano~I indicato
in figura~\ref{piani}.
      \end{fig}
\end{figure}

\clearpage

\begin{figure}
\includegraphics[width=13cm]{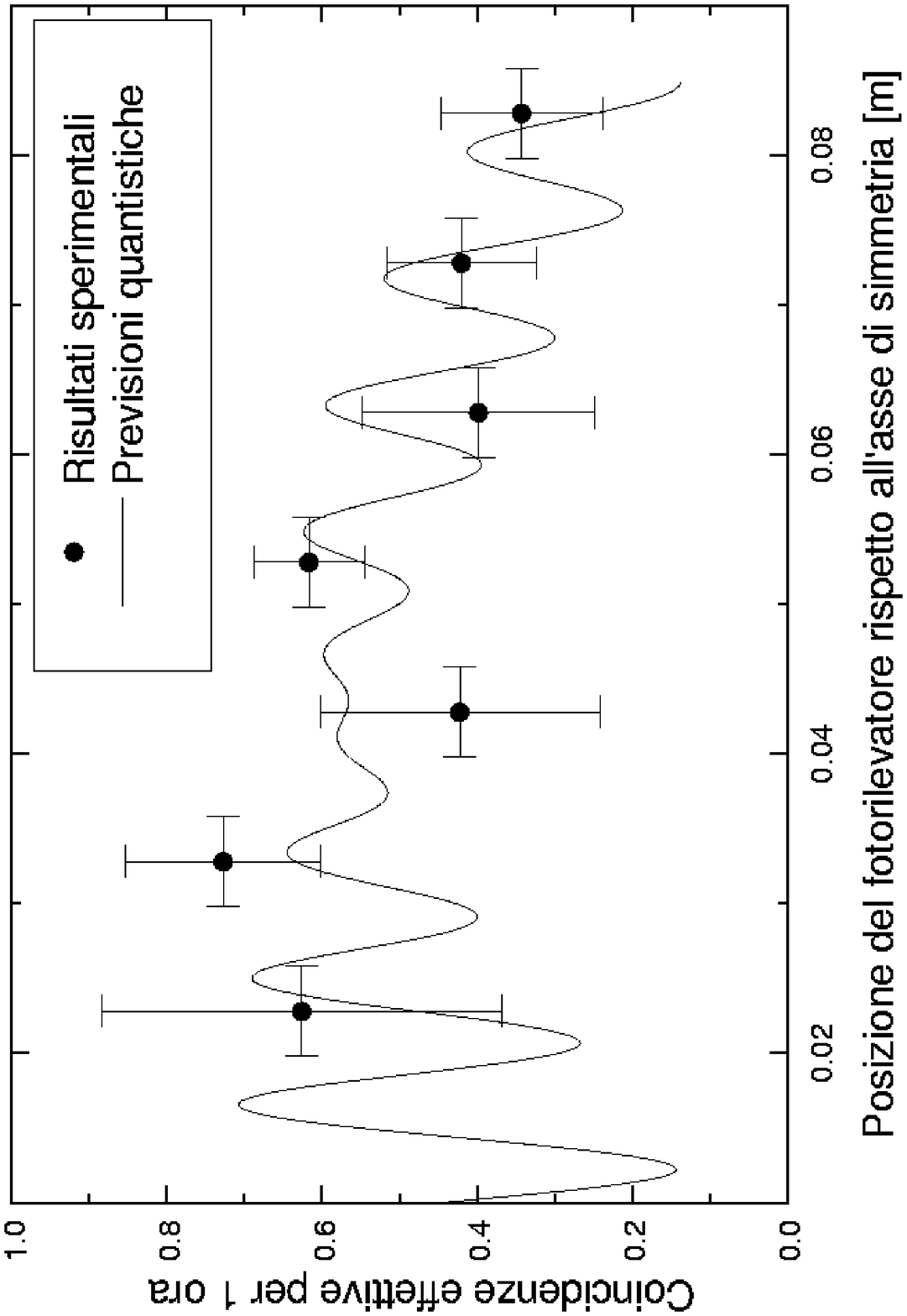}
      \begin{fig}\label{pattern2}
Profilo della figura di interferenza al quarto ordine 
prodotta 
dal passaggio di due fotoni entangled, 
di lunghezza d'onda 702~$nm$, 
attraverso una doppia fenditura,
ognuno per una determinata apertura.
Tale profilo può essere ricavato
dalla figura~\ref{graficomodulopsi3d}
sezionando con il piano~II indicato
in figura~\ref{piani}.\\
La configurazione sperimentale non consente
ancora una valutazione dettagliata 
delle frange di interferenza.
      \end{fig}
\end{figure}

\clearpage

\begin{figure}
\includegraphics[width=13cm]{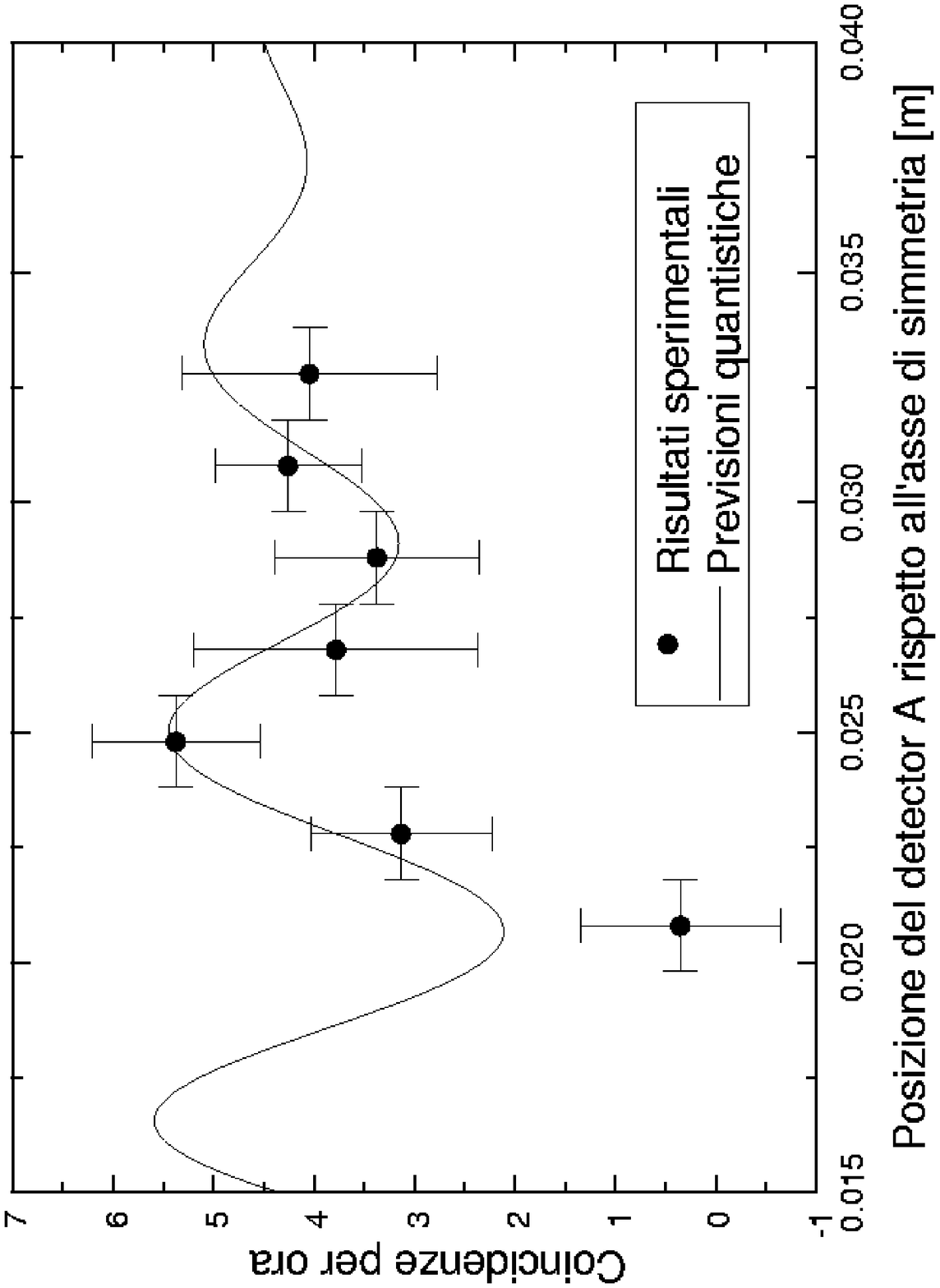}
      \begin{fig}\label{pattern3}
Profilo della figura di interferenza al quarto ordine 
prodotta 
dal passaggio di due fotoni entangled, 
di lunghezza d'onda 702~$nm$, 
attraverso una doppia fenditura,
ognuno per una determinata apertura.
Tale profilo può essere ricavato
dalla figura~\ref{graficomodulopsi3d}
sezionando con il piano~II indicato
in figura~\ref{piani}.\\
La configurazione sperimentale consente
una valutazione sufficientemente dettagliata 
delle frange di interferenza.
      \end{fig}
\end{figure}

\clearpage

\clearpage

\chapter{Confronto fra dBB e SQM}\label{dBB}

In meccanica quantistica 
ad una particella è associata una funzione
d'onda~$\psi$ che fornisce una descrizione 
completa del sistema la cui evoluzione temporale,
in una trattazione non-relativistica,
è data dall'
{\enfatizza equazione di
Schr\"odinger}:

\begin{equation}\label{equazionedisc}
i \hbar \frac{\partial \psi}{\partial t}%
=\frac{\hbar ^2}{2m} \nabla^2 \psi+V \psi
\end{equation}

Poichè tale equazione è lineare ed omogenea
le soluzioni formano uno spazio vettoriale lineare,
quindi normalizzabili, e soddisfano il 
{\enfatizza principio di sovrapposizione}: se $\psi_1$
e $\psi_2$ sono soluzioni lo è anche una 
loro combinazione lineare:
$\psi=\alpha \psi_1+\beta \psi_2$.

Nell'interpretazione della scuola di Copenhagen 
$\psi(t,\vec x)$ è una ampiezza di probabilità
da associarsi ad ogni singolo sistema fisico.\\
Le probabilità quantistiche non sono, quindi,
quelle tipiche di un sistema statistico,
ma sono legate alla natura intrinsecamente
stocastica della teoria.\\
Qualora si effettui una misura 
la probabilità di trovare,
all'istante $t$ il sistema 
nell'elemento di volume $dV$,
è determinata da 
$\int_V|\psi(t,\vec x)|^2 dV$,
dove $|\psi(t,\vec x)|^2$
è la densità di probabilità di posizione.\\
All'atto della misurazione si assiste,
quindi, al 
{\enfatizza
`collasso della funzione d'onda'
}\footnote{Si parla anche di riduzione 
del pacchetto d'onda del quanto.}, 
ovvero al meccanismo per cui di
tutte le soluzioni 
possibili 
solo una si realizza:
quella corrispondente al risultato misurato.
Tutto ciò, però,
non accade con una modalità deterministica,
ma stocastica:
prima della misura il sistema 
ha solo una data probabilità 
che la funzione d'onda che lo descrive
collassi in
uno degli autostati dell'operatore associato
all'osservabile fisica oggetto di studio
(data dall'autovalore corrispondente).\\
Il principio del collasso della funzione d'onda, 
però, 
non è descritto all'interno del formalismo
matematico della teoria e
non è universalmente accettato
nella comunità dei fisici a causa dei problemi 
concettuali che esso comporta.

Sorge, quindi, un dubbio di carattere
interpretativo sulla completezza della teoria:
la meccanica
quantistica descrive i fenomeni in termini statistici
perché la natura è intrinsecamente aleatoria
oppure
essa è il limite
stocastico di una teoria deterministica;
tale secondo punto di vista fu condiviso da
Einstein con l'affermazione: 
{\enfatizza `Dio non gioca a dadi con l'Universo'}.

Nel 1935 Einstein, Podolsky e Rosen, 
si posero il problema se la meccanica quantistica
potesse essere considerata una teoria completa,
a tal fine definirono
il concetto di elemento di realtà 
nel modo seguente:
{\parla `Se, senza perturbare in alcun modo il sistema,
è possibile predirre senza alcuna incertezza
il valore di una quantità fisica,
allora esiste un elemento di realtà
fisica in corrispondenza di tale quantità'
\cite{dispensemarco}}.

A tale definizione è legato un esperimento 
ideale noto come 
{\enfatizza `il paradosso EPR'}
dalle iniziali dei tre collaboratori.\\
Al fine di dare un esempio specifico del loro
ragionamento si espone la formulazione più
semplice, ma equivalente, data da Bhom.\\
Si consideri uno stato di singoletto
di due particelle di spin $1/2$:

\begin{equation}
|\psi\rangle =\frac{|\uparrow\rangle |\downarrow\rangle -|\downarrow\rangle |\uparrow\rangle }{\sqrt{2}}
\end{equation}

Dove $|\uparrow\rangle$ e $|\downarrow \rangle$ rappresentano
una singola particella di spin orientato
rispettivamente verso l'alto o verso il basso
lungo l'asse $z$ fissato ad arbitrio.\\
Si supponga che le due particelle siano separate
spazialmente e si misuri la componente~$z$
dello spin della seconda particella.
Quindi, la componente~$z$ della seconda particella
è un elemento di realtà secondo la definizione
precedente.\\
Si noti che lo stato di singoletto è invariante per
rotazione rispetto l'asse~$z$ (analoghe considerazioni valgono per
ogni altro asse), si può dunque concludere che ogni
altra componente di spin della seconda particella 
è un elemento di realtà.
Secondo la SQM, tuttavia, le componenti di spin
lungo assi differenti sono osservabili incompatibili,
alle quali non si può assegnare un valore definito
allo stesso tempo:
per tale motivo, secondo il ragionamento di
Einstein, Podolsky e Rosen, segue che la MQS
non può essere una teoria completa, in quanto
non permette una predizione del valore di ogni 
elemento di realtà.

Dall'analisi dell'impossibilità
di misurare contemporaneamente
elementi di realtà distinti
in meccanica quantistica 
Einstein, Podolsky e Rosen
conclusero che questa non è una 
teoria completa, ma l'approssimazione
di una teoria deterministica
ove tutte le osservabili hanno valori fissati
da variabili nascoste,
ponendo, quindi, le basi di 
quelle che sono note come teorie a variabili nascoste
(indicate brevemente con 
{\enfatizza HVT}) che possono essere locali
e non.

Nel 1964 il fisico irlandese John Stewart Bell ricava delle 
{\enfatizza disuguaglianze}
(teorema 
{\enfatizza di Bell}) che sono violate 
dalla meccanica quantistica
e non dalle teorie a variabili
nascoste {\sottolinea locali}, offrendo un criterio 
per distinguere sperimentalmente le due teorie.

Negli anni `80 presso i laboratori di Orsay a Parigi,
un gruppo di fisici guidati da Alain Aspect 
realizzò degli esperimenti,
basati sulle disuguaglianze di Bell,
che dimostrarono
l'indeterminazione intrinseca della natura
\footnote{
Già nel 1932 Johannes von Neumann
aveva escluso l'idea di teorie a variabili nascoste,
ma le sue conclusioni nascevano da ipotesi 
eccessivamente restrittive ed è per questo
che tali possibilità venne successivamente
ripresa in considerazione.}.

Tale lavoro portò all'eliminazione dell'ipotesi
di {\sottolinea teorie a variabili nascoste locali}\footnote{
Salvo l'ipotesi aggiuntiva di aver misurato un
campione fedele di quello complessivo, 
necessaria a causa della bassa efficienza di
rilevazione.}.
Esistono, però, teorie a variabili nascoste non locali,
per le quali le disuguaglianze di Bell
non valgono e che possono essere oggetto di studio,
la {\enfatizza teoria di de Broglie-Bhom},
oggetto di verifica della presente tesi,
ne è un esempio interessante.\\
La principale difficoltà che si incontra
nel realizzare un esperimento al fine di
distinguere la MQS dalle HVT
è dovuto al fatto che le teorie alternative
alla meccanica quantistica, costruite
per spiegare in modo differente
la natura dei fenomeni fisici,
devono riprodurre gli stessi risultati
sperimentali e quindi risultano indistinguibili
all'atto pratico.
Tuttavia, un recente lavoro condotto da un gruppo di
fisici indiani,
guidati dal prof. P.Ghose, prevede
delle condizioni in cui la 
Meccanica Quantistica Standard 
(nel seguito SQM) e la 
teoria di de Broglie~-~Bhom (dBB)
conducono a risultati diversi.\\
Prima di entrare nel dettaglio di
tale proposta occorre fornire alcuni 
cenni sulla teoria di de~Broglie-Bhom.

\section{Teoria di de~Broglie-Bhom}

La duplice natura ondulatoria e corpuscolare
della natura, per de~Broglie è concreta
e non solo un modo di presentarsi
dei fenomeni: un'onda 
{\sottolinea reale} accompagna davvero
il moto della particella e la guida,
questo è il concetto di 
{\enfatizza onda pilota}.
A queste conclusioni de Broglie arriva 
enfatizzando il limite classico dell'equazione
di Schr\"odinger, nel tentativo di
conservare il determinismo delle leggi
fisiche accanto ad una interpretazione 
statistica emergente, ma tale
idea viene bloccata sul nascere,
principalmente per le critiche di Pauli.
Al Quinto Congresso di Solvay,
tenutosi a Bruxelles dal 24 al 29
ottobre 1927, si sancisce
l'interpretazione stocastica
delle soluzioni dell'equazione di
Schr\"odinger.\\
L'idea di una teoria a variabili nascoste,
inizialmente introdotta 
da Einstein, Podolsky e Rosen,
criticata da Johannes von Neumann
viene successivamente ripresa da David Bhom
nel 1951.
Egli, pur ammettendo che la tradizionale
visione della SQM sia coerente,
non vuole escludere la possibilità che 
esistano altre interpretazioni,
che potrebbero essere
in grado di recuperare, in linea di principio,
una descrizione causale di tutti i processi
fisici.\\
Quando, nel 1951, David Bhom riprende il concetto
di onda guida, introdotta da de Broglie, 
sviluppa la sua teoria 
in un contesto in cui
per i fermioni si ha una concezione 
corpuscolare, mentre per i bosoni si adotta
una descrizione ondulatoria.\\
Questa teoria 
è un particolare esempio di HVT non-locale
dove la variabile nascosta è la posizione della particella.\\
Seguendo una procedura analoga a quella dell'approssimazione
semiclassica del metodo 
{\enfatizza WKB} 
si considera una funzione
d'onda descritta  in forma polare nel seguente modo:\\

\begin{equation}
\psi(\vec{r},t)=R(\vec{r},t)\cdot e^{\frac{i}{h}S(\vec{r},t)}
\end{equation}

Inserita nell'equazione di Schrodinger:

\begin{equation}
i\hbar \frac{\partial \Psi \left( \vec{r};t\right) }{\partial t}=%
\left[ -\frac{\hbar ^{2}}{2m}\triangle +V\left( \vec{r}\right) %
\right] \Psi \left( \vec{r};t\right)  \label{eq.Schrodinger}
\end{equation}

dove V è il potenziale classico,
separando parte reale e parte immaginaria,
si ottengono le seguenti due equazioni:

\clearpage

\begin{eqnarray}
\label{H-J}
\frac{\partial S}{\partial t}+\frac{\left( \nabla S\right) ^{2}}{2m}+V+Q=0\nonumber\\
\mbox{(eq. di diffusione nella forma di eq. di H.J.)}
\end{eqnarray}

\begin{eqnarray}
\label{continuità}
\frac{\partial \rho }{\partial t}+\frac{1}{m}\nabla \left( \rho
\vec{v}\right) =0 \nonumber\\
\mbox{(eq. di continuità)}
\end{eqnarray}

dove si è posto:

\begin{equation}
Q=-\frac{\hbar ^{2}}{2m}\frac{\triangle R}{R}
\end{equation}

\begin{equation}
\rho =\left| \Psi \right| ^{2}=R^{2}
\end{equation}

\begin{equation}
\vec{v}=\nabla S  \label{nomenclatura}
\end{equation}

Se V è il potenziale classico,
Q può essere considerato come un potenziale
quanto-meccanico che dipende
dall'ampiezza R della funzione d'onda $\psi$,
da cui segue il carattere non locale 
della teoria:
una variazione del potenziale in un 
punto dello spazio modifica 
istantaneamente la funzione d'onda
che descrive l'intero sistema.\\
Sono questi due potenziali che determinano
la traiettoria delle particelle 
secondo la legge di evoluzione:

\begin{equation}
\label{leggedievoluzione}
\frac{d^{2}\vec{r}}{dt^{2}}=-\nabla (V+Q)
\end{equation}

\begin{figure}[h]
%\begin{center}
\includegraphics[width=16cm]{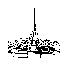}
      \begin{fig}\label{potenziale}
Potenziale quantistico associato ad una doppia fenditura.
      \end{fig}
%\end{center}
\end{figure}

\begin{figure}[h]
\begin{center}
\includegraphics[width=15cm]{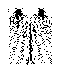}
      \begin{fig}\label{traiettorie2}
Traiettorie  seguite dalle particelle
che attraversano una doppia fenditura
secondo la meccanica di Bhom.
      \end{fig}
\end{center}
\end{figure}

Noto il potenziale quanto-meccanico
è possibile,
risolvendo l'eq. \ref{leggedievoluzione},
graficare  le traiettorie seguite dalle particelle.\\
Lo stesso Bhom propose un esperimento,
al fine di applicare il formalismo introdotto,
basato sulla tecnica dell'interferometro 
di Young.
Si consideri il caso di una doppia fenditura,
la quale è descritta dal potenziale 
riportato in fig.~\ref{potenziale}.
Si faccia l'ipotesi di utilizzare particelle
classicamente caratterizzate
da una posizione ben definita.\\
Per calcolare le traiettorie è sufficiente
risolvere l'eq. \ref{leggedievoluzione}
col potenziale considerato
(fig.~\ref{potenziale}).\\
Se la distribuzione delle posizioni
iniziali delle particelle è
data da \\$|\psi(x,t=0)|^2$,
tale risultato riproduce quello
della MQS per una popolazione di
particelle descritte dalla
funzione d'onda iniziale~$\psi(x,t=0)$.\\
Il risultato è riportato 
in fig.~\ref{traiettorie2}.\\
Nella versione iniziale della dBB fermioni
e bosoni erano dunque descritti in maniere diverse,
i primi come particelle dotate di traiettoria
i secondo come campi classici.
Recentemente si è dimostrato che, 
utilizzando il formalismo di
Kemmer-Duffin-Harishchandra~\cite{formalismoghose},
è possibile costruire una meccanica quantistica
relativistica per bosoni che 
conservi una quadri-corrente di probabilità
con la componente temporale definita positiva.
Ciò permette di definire delle traiettorie
anche per tali particelle.\\
Si può quindi tornare ad una descrizione `unificata'
di bosoni e fermioni anche nello schema dBB.\\
Come accennato precedentemente,
è stata avanzata, recentemente, la proposta 
di un possibile confronto tra la MQ e la 
dBB~\cite{propostaesperimentoghose},
che descriverò nel paragrafo seguente.

\clearpage

\clearpage

\clearpage

\section{Differenze tra dBB e SQM}

In questo capitolo descriverò la proposta di P.Ghose
per effettuare  un confronto sperimentale tra la dBB e la SQM.
Per tale fine è necessario 
introdurre
alcune nozioni sui sistemi dinamici.

\subsection{Ergodicità in SQM e dBB}

Il carattere non ergodico di un 
sistema dinamico è strettamente legato
alla differenza che esiste tra
media spaziale e media temporale.\\
La media spaziale di una funzione F
a valori complessi su M è così definita:

\begin{equation}
\begin{array}{lr}
\bar F= \int_M \rho(q,p) F(q,p)dqdp &, \int \rho(q,p)dqdp=1  
\end{array}
\end{equation}

La media temporale è data da:

\begin{equation}
F^{\star}=\lim_{N\rightarrow \infty}%
\frac{1}{N} \sum^{N-1}_{n=0} F(\phi^n_t q)
\end{equation}

Dove q e p indicano l'insieme di coordinate
e impulso, $\rho(q,p)dqdp$ è la misura 
definita nello spazio delle fasi
e $\phi_t:M \rightarrow N$ è una mappa
ad un parametro (il tempo)
consistente in un gruppo di misure che
preservano il diffeomorfismo.

Nell'ambito della teoria dei sistemi dinamici
esiste un teorema che mette in relazione
l'ergodicità con le medie 
precedentemente definite.

\begin{Teorema}\label{teoremaergodico}
{\parla `Se un sistema dinamico è ergodico
allora le medie spaziali e temporali,
di ogni funzione F a valori complessi 
su M, esistono e sono identiche;
se il sistema dinamico è non ergodico
ciò non può accadere.'}
\end{Teorema}

Partendo da tali definizioni, 
nel lavoro teorico di ref.\cite{formalismoghose}
si dimostrano i seguenti punti che
conducono ad una condizione che 
distingue la MQS dalla dBB.

\begin{enumerate}
\item Tutte le HVT, quindi anche la dBB,
sono costruite in modo da riprodurre 
i valori medi spaziali $\vec F$ dell'osservabile~F
come calcoli in SQM

\begin{equation}
\bar F (dBB) \equiv \bar F (SQM)
\end{equation}

\item Si può dimostrare che in SQM
tutti i sistemi dinamici sono 
ergodici quindi 
sulla base del teorema \ref{teoremaergodico}:
$\bar F (SQM) \equiv F^{\star} (SQM)$.

\item Nella dBB, invece,
esistono sistemi per i quali:
$\bar F (dBB) \neq F^{\star} (dBB)$,
ovvero sono non ergodici.

\item Dalle precedenti tre relazioni segue che:
$F^{\star}(dBB) \neq F^{\star}(SQM)$.

\end{enumerate}

In sintesi si può concludere che:

\begin{displaymath}
\begin{array}{ccr}
{\bar F}(dBB) \equiv       & {\bar F}(SQM)       & (1)  \\
\nparallel      (3)        & \parallel (2)       &      \\
F^{\star}(dBB) \neq        & F^{\star}(SQM)      & (4) 
\end{array}
\end{displaymath}

Queste relazioni implicano che 
ponendosi in particolari condizioni
(ed effettuando medie temporali e
non spaziali) si possono osservare 
differenze fra le due teorie
prese in esame,
come espresso dalla relazione~(4).
Ciò ha condotto alla 
proposta~\cite{propostaesperimentoghose}
di uno schema sperimentale
(descritto nel paragrafo successivo),
che è stato realizzato nell'ambito
della presente tesi.

\clearpage

\subsection{Esperimento proposto}

\begin{figure}[h]
\begin{center}
\includegraphics[width=15cm]{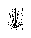}
      \begin{fig}\label{traiettorie3}
Traiettorie previste dalla meccanica di Bhom 
secondo ref.\cite{propostaesperimentoghose}
seguite dalle particelle 
che attraversano una doppia fenditura,
ognuna in una specifica apertura~(1 e 2).
      \end{fig}
\end{center}
\end{figure}

Con riferimento alla figura~\ref{traiettorie3} e
secondo i calcoli svolti in ref.~\cite{propostaesperimentoghose},
essendo i fotoni (1 e 2), che attraversano la doppia fenditura,
particelle è possibile 
calcolare le loro velocità: $\vec v_1$ e $\vec v_2$.
In particolare, ciò che interessa è la somma vettoriale: 
$\vec v_1+\vec v_2$.
I quanti utilizzati per il nostro esperimento sono 
prodotti mediante fluorescenza parametrica
e quindi 
(indicando con~$x$ di simmetria della doppia fenditura
parallelo al fascio di pompa) si ha:

\begin{equation}\label{p}
\vec v_{1y}+\vec v_{2y}=0
\end{equation}

Dalla \ref{p} (con riferimento alla figura~\ref{traiettorie3})
segue che per ogni istante~$t$ si ha:

\begin{equation}
y_1(t)+y_2(t)=y_1(0)+y_2(0)
\end{equation}

Se le posizioni iniziali delle due particelle sono simmetriche
rispetto l'asse~$x$ (cioè $y_1(0)+y_2(0)=0$),
allora esse non attraverseranno mai l'asse di simmetria
poichè:

\begin{equation}
y_1(t)=-y_2(t)
\end{equation}

Nella nostra esperienza i fotoni che attraversano
la doppia fenditura, ognuno per una specifica fenditura, sono entangled
e quindi il loro percorso può essere individuato 
mediante l'uso dell'apparato di rilevazione 
descritto nel paragrafo~\ref{apparatodirilevazione}.
Posizionando entrambi i fotorivelatori a sinistra
o adestra dell'asse di simmetria~$y=0$ in fig.~\ref{traiettorie3}
secondo le predizioni della dBB di 
ref.~\cite{propostaesperimentoghose} non dovrebbero 
rilevarsi coincidenze in quanto tale evento
implicherebbe che uno dei fotoni
della coppia entangled abbia attraversato l'asse
di simmetria della doppia fenditura.

La MQS, invece, prevede un numero di coincidenze non nullo
in questa configurazione, come illustrato precedentemente
(la figura di interferenza al quarto ordine
dovuta alle coincidenze non si annulla).

\clearpage

\section{Risultati sperimentali}

Durante la prima scansione della figura di 
diffrazione si è raggiunta una configurazione in 
cui (con riferimento alla figura~\ref{schemaasse})
il fotorivelatore~A si trova a~-5.5~cm
il~B a~-1.7~cm~\footnote{
Le distanze rispetto l'asse di simmetria 
sono riferite all'asse ottico delle
lenti montate sul fotorivelatore.
}.
I due apparati di rilevazione sono
entrambi nello stesso semipiano,
quindi, per~\cite{propostaesperimentoghose} e~\cite{formalismoghose}
non dovrebbero essere rilevate coincidenze
per la MQS ciò, invece, è possibile.
Dopo 35 acquisizioni da 30 minuti ciascuna
si sono ottenuti $78\pm10$ coincidenze:
tale risultato è perfettamente compatibile
con la MQS, ma è in contrasto con le predizioni
di~\cite{propostaesperimentoghose} per la dBB
di quasi 8~deviazioni standard.

\begin{figure}[h]
\begin{center}
\includegraphics[width=14cm]{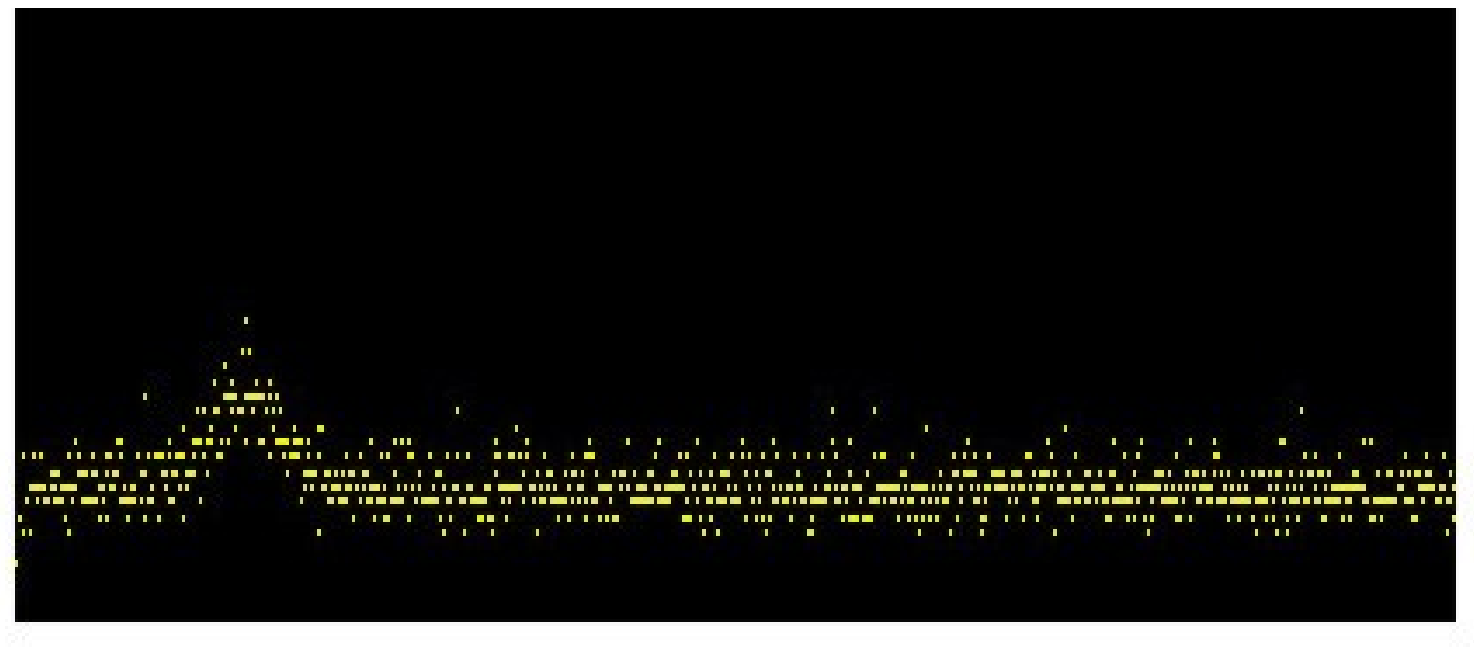}
      \begin{fig}\label{picco2}
Immagine del picco di coincidenze restituito dal
multicanale con i fotorivelatori a -5.5~cm 
il sinistro e -1.7~cm il destro.
In questa configurazione, nella teoria di dBB secondo i
calcoli di P.Ghose~\cite{propostaesperimentoghose}
non si dovrebbe osservare nessun picco.
      \end{fig}
\end{center}
\end{figure}

La MQS prevede anche un picco secondario
di coincidenze smpre dentro
alla regione ove la dBB prevede nessuna coincidenza,
ma con i fotorivelatori ancora
più distanti dall'asse di simmetria
della doppia fenditura.
Tale picco è stato misurato posizionando il
il fotorivelatore~A a~-11.7~cm
ed il~B a~-4.4~cm.
In questa posizione si sono registrate $41\pm14$
coincidenze, dopo 17 acquisizioni da 1 ora.\\
Questi dati sono stati raccolti posizionando
i rilevatori sempre nella regione in cui 
secondo la ref.~\cite{propostaesperimentoghose}
le coincidenze rilevabili dovrebbero
essere nulle.\\
Essi sono quindi perfettamente compatibili
con la SQM, ma pongono uno stringente
limite sulla validità della teoria dBB
alla base dei calcoli di ref.\cite{propostaesperimentoghose}.

\section{Considerazioni sull'esperimento}\label{considerazioni}

I risultati ottenuti nel nostro esperimento,
riguardante il confronto fra dBB e SQM, hanno suscitato
un certo clamore nel mondo della ricerca scientifica.

Ad esempio tale lavoro è stato oggetto di un
articolo su una rivista divulgativa~\cite{articolonewscientist}.
Numerosi commenti sono poi apparsi, alcuni di 
essi contenenti dubbi sulla validità della proposta 
di ref.~\cite{propostaesperimentoghose}.\\
Una prima critica 
è stata pronunciata da Antony Valentini
(Imperial College, Londra),
i cui dubbi
riguardano la non certezza 
sulla possibilità che 
i fotoni (che attraversano le due fenditure simultaneamente)
siano effettivamente prodotti nello
stesso punto del cristallo.
A tale osservazione è stato risposto
evidenziando il fatto che
per costruzione (vedi capitolo~\ref{apparatodirilevazione})
l'esperimento seleziona solo coppie 
i cui fotoni costituenti siano indirizzati
verso una ben precisa fenditura.
Coppie prodotte in altre zone del cristallo
non verificano tale condizione e
contribuiscono alla formazione di rumore di
fondo (che può essere stimato, come illustrato nel paragrafo~\ref{paragrafotaratura}) durante l'acquisizione dati.\\
Più genericamente alcuni  altri sostenitori
della teoria dBB~\cite{articolobhom} hanno affermato che le 
due teorie (SQM e dBB) devono essere
equivalenti `per costruzione' e quindi dev'esservi
un errore nella proposta teorica di ref.\cite{propostaesperimentoghose}.
Tuttavia tale errore non è stato sinora identificato.\\
In conclusione, i risultati del nostro esperimento
confermano pertanto la SQM e contraddicono
i risultati della dBB
(almeno nella versione in cui i fotoni siano
trattati come particelle)
di ref.\cite{propostaesperimentoghose},
sollecitando una chiarificazione definitiva 
della validità della meccanica quantistica standard.

\begin{figure}[h]
\begin{center}
\includegraphics[width=13cm]{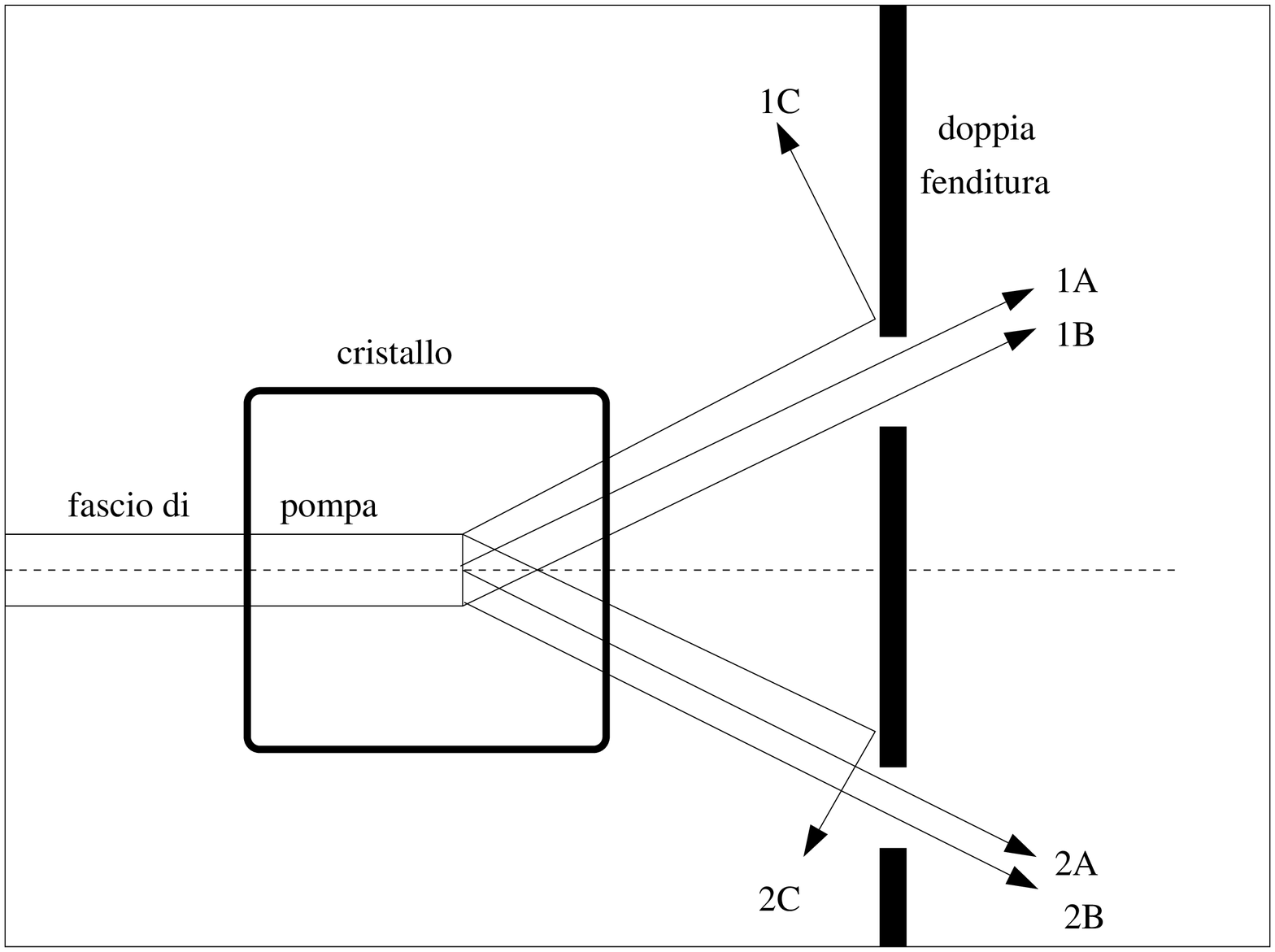}
      \begin{fig}\label{laserincristallo}
Geometria delle traiettorie delle coppie di fotoni
correlati emessi per fluorescenza parametrica.\\
Iniettando un fascio di pompa nel cristallo $LiO_3$
si generano, per fluorescenza parametrica,
coppie di fotoni entangled;
il diametro di tale fascio, però,
non è nullo, a seconda delle ottiche utilizzate
può essere di 1mm o 2mm,
quindi la produzione di coppia non avviene in un punto
bensì in una regione spaziale finita.
      \end{fig}
\end{center}
\end{figure}

\clearpage

\appendix

\chapter{looptest.pas}\label{looptest.pas} 
 
\begin{verbatim} 
 
program C842UT4;          (* intestazione: consiste nel nome del programma   *) 
 
uses crt, qfl360 ;        (* indica la libreria cui fare riferimento, 
                             cioè: qfl360 *) 
var                       (* definisco le variabili:                         *) 
   rep   : string;        (* rep : è una stringa                             *) 
   i     : integer;       (* i     :è un intero ,lo uso per contare i loop   *) 
                          (* il campo di variabilità è: da -32768 a 32768    *) 
   status: integer;       (* status: è un intero                             *) 
   A     : longint;       (* A     : è un intero lungo                       *) 
                          (* variabilità è: da -2147483648 a 2147483647      *) 
   B     : longint;       (* B     : è un intero lungo                       *) 
   L     : longint;       (* L     : è un intero lungo                       *) 
   t     : real;          (* t     : è un numero reale                       *) 
   r     : real;          (* r     : è un numero reale                       *) 
 
begin 
   clrscr;                (* sta per clear screen: 
                             cancella lo sfondo quando compare il DOS        *) 
   Set_BaseAddress(210);  (* definisce la board address: 
                             dice al turbo pascal dove può mettere in memoria 
                             nella RAM i suoi dati 
                             in questo caso dall'indirizzo 210 in poi        *) 
 
   MaxAxis := Axis_Installed; (* sono comandi che fanno                      *) 
   status := InitBoard(9);    (* riferimento a indicazioni                   *) 
   {Init_LS(0,0,0,0);}        (* presenti nella libreria qfl360              *) 
 
   if autodetect = -1 then begin (* se il comando autodetect 
                                    definito nelle librerie, vale -1 allora  *) 
                                 (* scrivi questa frase                      *) 
      writeln('ERROR: at least one limit switch was hit'); 
      halt;                       (* e poi fermati                           *) 
   end; 
 
   InitAxis(1);                       (* sono comandi                        *) 
   InitAxis(2);                       (* che fanno riferimento               *) 
   if MaxAxis > 2 then begin          (* a indicazioni                       *) 
      InitAxis(3);                    (* presenti nella                      *) 
      InitAxis(4);                    (* libreria qfl360                     *) 
   end; 
 
(*****************************************************************************) 
(* Fare dei loop stoppati con indicazione della posizione  :                 *) 
(*****************************************************************************) 
 
execute(1,sv,176470,rep);     (* velocità di settaggio del traslatore        *) 
execute(2,sv,250000,rep);     (* velocità di settaggio del rotatore          *) 
 
for i:=1 to 2 do begin             (* imposto 2 loop                         *) 
 
writeln('STO TRASLANDO');      (* scrivi : " STO TRASLANDO "                 *) 
t :=10 ;                       (* passo della traslazione in mm              *) 
A :=round((17647059/150)*t);   (* trasforma t in conteggi                    *) 
 
execute(1,MR,A,rep);           (* muovi il motore n1 per A conteggi relativi *) 
execute(0,tp,0,rep);           (* leggi tutte le 4 posizioni                 *) 
execute(0,ws,0,rep);           (* aspetta che tutti i motori siano fermi     *) 
 
writeln('STO RUOTANDO');       (* scrivi : " STO RUOTANDO "                  *) 
L :=500;                       (* imposto la distanza contatore-sorgente 
                                         in mm*) 
r :=arctan(t/L) ;              (* calcola l'angolo di rotazione 
                                                 in funzione della posizione *) 
B :=round((10671111/(2*pi))*r);(* trasforma l'angolo di rotazione in conteggi*) 
execute(2,MR,B,rep);           (* muovi il motore n2 per B conteggi relativi *) 
execute(0,tp,0,rep);           (* leggi tutte le 4 posizioni                 *) 
execute(0,ws,0,rep);           (* aspetta che tutti i motori siano fermi     *) 
 
writeln('STO CONTANDO');       (* scrivi : " STO CONTANDO "                  *) 
delay(1500);                   (* aspetta 1500 frquenze di clock             *) 
                                       execute(0,ws,0,rep); 
                               (* aspetta che tutti i motori siano fermi     *) 
execute(0,tp,0,rep);           (* leggi tutte le 4 posizioni                 *) 
write(rep);                    (* scrivi le posizioni lette                  *) 
writeln(get_pos(2):12);        (* scrivi il valore letto del asse n2         *) 
 
end;                               (* fine del loop                          *) 
 
writeln('STO TORNANDO A ZERO');    (* scrivi : " STO TORNANDO A ZERO "       *) 
execute(0,MA,0,rep);               (* azzera tutti i motori                  *) 
execute(0,tp,0,rep);               (* leggi tutte le 4 posizioni             *) 
execute(0,ws,0,rep);               (* aspetta che tutti i motori siano fermi *) 
writeln('FINE');                   (* scrivi : " FINE "                      *) 
 
end.                               (* fine del programma looptest.pas        *) 
 
\end{verbatim} 
 
\chapter{quad.pas}\label{quad.pas}

\begin{verbatim} 
 
program M974; 
uses strings, wincrt, wintypes, winprocs; 
 
const 
     ComName: String = 'COM1'; 
     InQueue: Word = 128; 
     OutQueue: Word = 128; 
     ModeDef: String = 'COM1:9600,e,8,1'; 
 
var 
     Dcb, DcbChk : TDCB; 
     ComId : Integer; 
     Status : Integer; 
     (* Definizione di variabili Null Terminated Strings 
     da urilizzare 
     nelle funzioni BuilCommState e SetCommState*) 
     N_ComName : array[0..80] of Char; 
     N_ModeDef : array[0..80] of Char; 
     N_TxBuffer : array[0..80] of Char; 
     N_RxBuffer : array[0..80] of Char; 
     TxLen : Integer; 
     Command : String; 
     Answer : String; 
     StartTime : LongInt; 
 
begin 
 
(* Apertura/impostazione int. seriale *) 
StrPCopy(N_ComName, ComName); 
ComId := OpenComm(N_ComName, InQueue, OutQueue); 
if ComId < 0 then 
begin 
     writeln('OpenComm() fallita!'); 
     Exit 
end; 
writeln('OpenComm() OK: ', ComId); 
 
(* BuildCommDCB() imposta l'Id del Dcb al ComId 
   ricavato da N_ModeDef *) 
StrPCopy(N_ModeDef, ModeDef); 
Status := BuildCommDCB(N_ModeDef, Dcb); 
if Status < 0 then 
begin 
     writeln('BuildCommDCB() fallita'); 
     Exit 
end; 
writeln('BuildCommDCB() OK'); 
 
(* Solo per controllo *) 
writeln('  BaudRate=', Dcb.BaudRate); 
writeln('  ByteSize=', Dcb.ByteSize); 
 
Status := SetCommState(Dcb); 
if Status < 0 then 
begin 
     writeln('SetCommState() fallita!'); 
     Exit 
end; 
writeln('SetCommState() OK'); 
 
(* Solo per controllo *) 
Status := GetCommState(ComId, DcbChk); 
if Status < 0 then 
begin 
     writeln('GetCommState() fallita!'); 
     Exit 
end; 
writeln('GetCommState() OK'); 
writeln('  BaudRate=', DcbChk.BaudRate); 
writeln('  ByteSize=', DcbChk.ByteSize); 
 
repeat 
      write('Comando? ');  readln(Command); 
 
      (* Inserire READ/WRITE qui *) 
      StrPCopy(N_TxBuffer, Command + Chr(10)); 
      TxLen := StrLen(N_TxBuffer); 
 
      Status := WriteComm(ComId, N_TxBuffer, TxLen); 
      if Status <> TxLen then 
      begin 
           writeln('WriteComm() fallita, Status=', Status, 'TxLen=', TxLen);                                    Exit 
      end; 
 
      (* Attenzione che GetTickCount va in overflow 
         dopo 49 giorni circa dall'inizio di attività continuativa 
         del sistema operativo Windows, 
         ovvero dopo circa 4.233.600.000 millisecondi  *) 
 
      StartTime := GetTickCount; 
      Answer := ''; 
      repeat 
            Status := ReadComm(ComId, N_RxBuffer, 80); 
            if Status >= 0 then 
            begin 
                 N_RxBuffer[Status] := Chr(0); 
                 if Status > 0 then  Answer := Answer + StrPas(N_RxBuffer); 
            end 
      until GetTickCount - StartTime > 1000; 
      (* Il controllo x-y >  1000 NON va bene in caso di overflow *) 
 
      writeln('Risposta=', Answer); 
until False; 
 
(* Chiusura int. seriale *) 
Status := CloseComm(ComId); 
if Status < 0 then 
begin 
     writeln('CloseComm() fallita!'); 
     Exit 
end; 
writeln('CloseComm() OK'); 
end. 
 
\end{verbatim}

\chapter{Experimental realization of a first test of de~Broglie-Bhom theory.}\label{pubblicazione}

\chapter{An innovative biphotons double slit experiment.}\label{pre-preprint}

\chapter{Is the last hope for certainty gone?}\label{pubblicazionenewscientist}

\end{document}